\documentclass[12pt]{article}
\usepackage{amsmath,comment}
\usepackage{graphicx,psfrag,epsf,color}
\usepackage{enumerate}
\usepackage{natbib}
\usepackage{url} 

\newcommand{\blind}{0}

\newcommand{\pkg}[1]{{\normalfont\fontseries{b}\selectfont #1}} 
\let\proglang=\textit

\usepackage{mathtools,amssymb}
\usepackage{bm}
\usepackage[linesnumbered,ruled,vlined]{algorithm2e}
\let\oldnl\nl
\newcommand{\nonl}{\renewcommand{\nl}{\let\nl\oldnl}}

\usepackage{titlesec}
\titlelabel{\thetitle.\quad}
\titleformat{\section}
  {\normalfont\fontsize{15}{15}\bfseries}{\thesection.}{1em}{}
\titleformat{\subsection}
  {\normalfont\fontsize{13}{13}\bfseries}{\thesubsection.}{1em}{}

\newtheorem{lemma}{Lemma}
\newtheorem{proposition}{Proposition}

\newcommand{\bB}{\bm{B}}

\newcommand{\bD}{\bm{D}}
\newcommand{\bX}{\bm{X}}
\newcommand{\bI}{\bm{I}}
\newcommand{\bU}{\bm{U}}
\newcommand{\bV}{\bm{V}}

\newcommand{\bx}{\bm{x}}
\newcommand{\bb}{\bm{b}}

\newcommand{\bK}{\bm{K}}

\newcommand{\by}{\bm{y}}
\newcommand{\bY}{\bm{Y}}
\newcommand{\bc}{\bm{c}}
\newcommand{\bk}{\bm{k}}

\newcommand{\br}{\bm{r}}
\newcommand{\bv}{\bm{v}}

\newcommand{\balp}{\bm{\alpha}}
\newcommand{\bbt}{\bm{\beta}}
\newcommand{\btht}{\bm{\theta}}
\newcommand{\bTht}{\bm{\Theta}}
\newcommand{\hbt}{\hat{\bm{\theta}}}
\newcommand{\hbT}{\hat{\bm{\Theta}}}
\newcommand{\hsq}{\hat{\sigma}^2}

\newcommand{\tbk}{\tilde{\bm{k}}}
\newcommand{\tbK}{\tilde{\bm{K}}}
\newcommand{\tr}{\text{tr}}
\newcommand{\diag}{\text{diag}}
\newcommand{\tE}{\text{E}}
\newcommand{\hbc}{\hat{\bm{c}}}

\DeclarePairedDelimiter{\ceil}{\lceil}{\rceil}
\begin{document}

\def\spacingset#1{\renewcommand{\baselinestretch}%
{#1}\small\normalsize} \spacingset{1}


\if0\blind
{
  \title{\bf Local Gaussian Process Model for Large-scale Dynamic Computer
  Experiments}
  \author{Ru Zhang, C. Devon Lin\hspace{.2cm}\\
    Department of Mathematics and Statistics, \\
    Queen's University, ON, Canada\\
    and\\
    Pritam Ranjan\\
    Operations Management \& Quantitative Techniques, \\
    Indian Institute of Management Indore, MP, India}
  \maketitle
} \fi

\if1\blind
{
  \bigskip
  \bigskip
  \bigskip
  \begin{center}
    {\LARGE\bf  Local Gaussian Process Model for Large-scale Dynamic Computer
  Experiments}
\end{center}
  \medskip
} \fi

\bigskip
\begin{abstract}
The recent accelerated growth in the computing power has generated  popularization of experimentation with dynamic computer models in various physical and engineering applications. Despite the extensive statistical research in computer experiments, most of the focus had been on the theoretical and algorithmic innovations for the design and analysis of computer models with scalar responses.

In this paper, we propose a computationally efficient statistical emulator for a large-scale dynamic computer simulator (i.e., simulator which gives time series outputs). The main idea is to first find a good local neighborhood for every input location, and then emulate the simulator output via a singular value decomposition (SVD) based Gaussian process (GP) model. We develop a new design criterion for sequentially finding this local neighborhood set of training points.
Several test functions and a real-life application have been used to demonstrate the performance of the proposed approach over a naive method of choosing local neighborhood set using the Euclidean distance among design points.

The supplementary material, which contains proof of the theoretical results, detailed algorithms, additional simulation results and \proglang{R} codes, are available online. 
\end{abstract}

\noindent%
{\it Keywords:} Nearest neighbor; Sequential design;
Singular value decomposition; Statistical emulator; Time series output.
\vfill

\newpage
\spacingset{1.45} 

\section{Intoduction}
\label{sec:introduction}

Computer experiments are increasingly used in physical, engineering and social sciences as an economical alternative to physical experiments with complex systems/phenomena (\citet{sacks1989design}; \citet{santner2003design}). Such experiments are performed on computers with the underlying process represented and implemented by mathematical models. Although cheaper than physical experiments, realistic computer experiments for complex processes can still be time-consuming or sometimes infeasible, and thus, statistical surrogates or emulators are often used for thorough investigation.

Popular objectives of such computer experiments include estimation of pre-specified process features (e.g., overall response surface, global optimum, inverse problem, quantile, and so on), sensitivity analysis, calibration and uncertainty quantification (\citet{jones1998efficient}; \citet{kennedy2001bayesian};  \citet{ranjan2008sequential}; \citet{bingham2014_EIreview}). Despite the extensive statistical research in computer experiments, most of the focus had been on the theoretical and algorithmic innovations for the design and analysis of computer models with scalar responses. In this paper we focus on the emulation of dynamic computer models - referred to computer simulators with time series outputs.

Dynamic computer experiments arise in various applications, for
example, rainfall-runoff model (\citet{conti2009gaussian}), and vehicle
suspension system (\citet{bayarri2007computer}). Our motivating application comes from an apple farming industry where the objective is to emulate the population growth curve of European red mites which infest on apple leaves and diminish the crop quality (\citet{teismann2009}).

With the accelerated growth of computing power, and hence the availability of dynamic computer simulators, there is a desperate need for innovative methodologies and algorithms for the design and analysis of experiments that can particularly handle large data sets. In general, the size of data is a multiple of the length of the time series outputs.
Recently, a few
  attempts on the emulation of dynamic computer experiments have been made by considering time as another input variable in the
  correlation structure and emulating the response via GP models
  \citep{stein2005space,conti2010bayesian,hung2015analysis}.
\citet{conti2009gaussian}
constructed dynamic emulators by using a one-step transition function
of state vectors to emulate the computer model movement from one time
step to the next. \citet{liu2009dynamic} proposed time varying
autoregression (TVAR) models with GP residuals. \citet{farah2014bayesian} extends the TVAR models in \citet{liu2009dynamic} by including the  input-dependent dynamic regression term.
Another clever
approach is to represent the time series outputs as linear
combinations of a fixed set of basis such as singular vectors
(\citet{higdon2008computer}) or wavelet basis
(\citet{bayarri2007computer}) and impose GP models on the linear
coefficients. However, fitting GP models over the entire training set can often be computationally infeasible for large-scale dynamic computer experiments involving thousands of training points.

We propose a new approach based on  singular value decomposition (SVD) and the local surrogate idea, the latter of which was originally proposed for scalar valued computer simulators with large training data (\citet{emery2009kriging}). The local surrogate idea was to emulate the process in a local neighborhood of the input location of interest.
A naive method of searching for local neighborhood is to select data close to the input location for prediction such that the selected input locations are distributed as uniformly as possible around the location for prediction (as in \emph{k-nearest neighbors}). This method does not take the spatial correlation into account. To search for the most relevant data for local neighborhood in a more intelligent way,  \citet{emery2009kriging} built a local neighborhood by sequentially including data that make the kriging variance decrease more.  \citet{gramacy2015local} further improved the prediction accuracy by using a sequential greedy algorithm and an optimality criterion for finding a non-trivial local neighborhood set. Our objective is to generalize this optimality criterion for the sequential construction of the local neighborhood set for emulating the dynamic computer simulators. We also develop an algorithm for the implementation of the proposed methodology which is efficient from a large-data standpoint.

The subsequent sections are organized as follows. Section
\ref{sec:svd-gpm} reviews the concept of SVD-based GP models and
provides a rigorous account for its model assumption and empirical
Bayesian inference.  Section \ref{sec:loc-svdgp} presents an innovative generalization of the optimality criterion, and a new algorithm for the local approximate SVD-based GP models. We also compare the computational complexity of the algorithms. Section \ref{sec:simu-local} uses two test functions to compare the performance of the  $k$-nearest neighbor SVD-based GP models (Euclidean distance based nearest neighbor), the full SVD-based GP models using all training points, and the proposed methodology in terms of prediction accuracy. The proposed method is also applied to the two-delay blowfly (TDB) model which simulates the population growth curve of European red mites. The concluding remarks are provided in Section~\ref{sec:con-rmks}, and proofs are given in the Supplementary Materials.

\section{SVD-based GP Models}
\label{sec:svd-gpm}

\citet{higdon2008computer} proposed an SVD-based GP model for the
calibration of computer simulators with highly multivariate outputs.
They used a full Bayesian approach for model fitting which is exceedingly expensive for large-scale computer experiments, particularly in our proposed sequential procedure for fitting local SVD-based GPs.  Thus, we first present a brief review of the SVD-based GP models proposed by \citet{higdon2008computer}, and then outline an empirical Bayesian procedure to reduce the computational burden.

\subsection{Model Formulation}
\label{sec2.1}

Consider a computer simulator which takes a $q$-dimensional quantitative input $\bm{x}\in \mathbb{R}^q$, and returns a time series output $\bm{y}(\bm{x})\in\mathbb{R}^L$ of length $L$.

For $N$ training points, let $\bm{X}=[\bm{x}_1,\dots,\bm{x}_N]^T$ be the $N\times q$ input matrix and $\bm{Y}=[\bm{y}(\bm{x}_1),\dots,\bm{y}(\bm{x}_N)]$ be the $L\times N$ matrix of time series responses. The SVD on $\bm{Y}$ gives
\begin{align*}
\bm{Y}=\bm{U}\bm{D}\bm{V}^T,
\end{align*}
where $\bm{U}=[\bm{u}_1,\dots,\bm{u}_{k}]$ is an $L\times k$ column-orthogonal matrix of left singular vectors, with $k$ being the minimum of $N$ and $L$, $\bm{D}=\text{diag}(d_1,\dots,d_{k})$ is a $k \times k$ diagonal matrix of singular values sorted in decreasing order, and the matrix $\bm{V}$ is an $N\times k$ column-orthogonal matrix of right singular vectors. The SVD-based GP model assumes that, for any $\bx\in\mathbb{R}^q$,
\begin{align}\label{eq:md}
\bm{y}(\bm{x})=\sum_{i=1}^pc_i(\bm{x})\bm{b}_i+\bm{\epsilon},
\end{align}
where the orthogonal basis $\bm{b}_i=d_i\bm{u}_i\in \mathbb{R}^L$, for $i=1,\dots,p$, are the first $p$ vectors of $\bm{U}$ scaled by the corresponding singular values.
The coefficients $c_i$'s in (\ref{eq:md}) are random functions
(\citet{rasmussen2006gaussian}) assumed to be independent Gaussian processes, i.e., $c_i\sim \text{GP}(0,\sigma^2_iK_i(\cdot,\cdot;\btht_i))$ for
$i=1,\dots,p$. We use the popular anisotropic Gaussian correlation,
\begin{align*}
K(\bm{x}_1,\bm{x}_2;\btht)=\exp\left\{-\sum_{j=1}^q\theta_j(x_{1j}-x_{2j})^2\right\},
\end{align*}
for characterizing the spatial correlation structure, however, one can easily use another suitable correlation structure like Mat\'ern or power-exponential (see \citet{santner2003design, rasmussen2006gaussian}). The residual error $\bm{\epsilon}$ in (\ref{eq:md})  is assumed to be independent Gaussian white noise, that is,  $\bm{\epsilon} \sim\mathcal{N}(0,\sigma^2\bI_L)$.
For notational simplicity, we denote
$\bm{U}^*=[\bm{u}_1,\dots,\bm{u}_p]$,
$\bm{D}^*=\text{diag}(d_1,\dots,d_p)$,
$\bm{V}^*=[\bm{v}_1,\dots,\bm{v}_p]$ and
$\bm{B}=[\bm{b}_1,\dots,\bm{b}_p] = \bm{U}^*\bm{D}^*$. The $j$th entry ($1\le j \le N$) of the $N$-dimensional vector $\bv_i$  ($1\le i \le p$) can also be interpreted as a realization of the Gaussian process model for $c_i(\bx_j)$.


The number of significant singular values, $p$ in (\ref{eq:md}), is determined empirically by the cumulative percentage criterion
\begin{equation}\label{eq:pp}
p=\min\left\{m:\frac{\sum_{i=1}^md_i}{\sum_{i=1}^{k}d_i}>\gamma\right\},
\end{equation}
where $\gamma$ is a prespecified threshold of explained variation (we used $\gamma=0.95$).

Similar to \citet{higdon2008computer}, we use Bayesian algorithms for model fitting, however, since a full Bayesian implementation is too time consuming, we follow an empirical Bayesian approach. This is particularly crucial here as the GP models have to be fit several times in the proposed sequential procedure.

\subsection{Empirical Bayesian Inference}

This section briefly reviews the key components of our model fitting procedure.
For all the model parameters, we use the maximum a posteriori (MAP) values as the plug-in estimates. The parameters of interest are $\sigma^2$ - the error variance, and for $i =1, 2,...,p$, the process variance $\sigma^2_i$ and the $q$-dimensional correlation hyper-parameter $\btht_i = (
\theta_{i1},\ldots, \theta_{iq})$. Similar to  \citet{gramacy2015local}, we use inverse Gamma priors for $\sigma^2_i$ and $\sigma^2$, i.e.,
\begin{align*}
  \begin{aligned}
    [\sigma_i^2]\sim\text{IG}\left(\frac{\alpha_i}{2},\frac{\beta_i}{2}\right),
    i=1,\dots,p,
  \end{aligned} \qquad
  \begin{aligned} [\sigma^2]\sim \text{IG}\left(\frac{\alpha}{2},\frac{\beta}{2}\right),
  \end{aligned}
\end{align*}
and use the Gamma prior for the hyper-parameter $1/\theta_{ij}$ with the shape parameter $3/2$ and the scale parameter chosen such that the maximum squared distance among any two points of the design matrix lies at the position of $95\%$ quantile (\citet{gramacy2015lagp}). As a result, the posterior of $\btht_i$ becomes

\begin{equation}\label{eq:post}
\pi(\bm{\theta}_i|\bv_i)\propto|\bm{K}_i|^{-\frac{1}{2}}\left(\frac{\beta_i+\psi_i}{2}\right)^{-(\alpha_i+N)/2}\pi(\bm{\theta}_i),
\end{equation}

\noindent where $\pi(\bm{\theta}_i)$ represents the prior of $\bm{\theta}_i$, $\bm{K}_i$ is the $N\times N$ correlation matrix
on the training set $\bm{X}$ with the $(j,k)$th entry being
$K(\bm{x}_j,\bm{x}_k;\btht_i)$, for $j,k=1,\dots,N$,
\begin{align*}
  \psi_i=\bv_i^T\bm{K}_i^{-1}\bv_i,
\end{align*}
\noindent and $\bv_i$ is the $i$th column of $\bm{V}^*$.


It can also be shown that, for any input $\bx_0$, the conditional distribution of $c_i(\bx_0)$ given $(\bv_i,\bm{\theta}_i)$ is independent non-central $t$ distribution with $N+\alpha_i$ degrees of freedom, for $i=1,\dots,p$, i.e.,
\begin{align*}
[c_i(\bm{x}_0)|\bv_i,\bm{\theta}_i]\sim
t_{N+\alpha_i}\left(\hat{c}_i(\bm{x}_0|\bv_i,\bm{\theta}_i),\hat{\sigma}_i^2(\bm{x}_0|\bv_i,\bm{\theta}_i)\right),
\end{align*}
where the location parameter is
\begin{align*}
\hat{c}_i(\bm{x}_0|\bv_i,\bm{\theta}_i)=\bm{k}_i^T(\bm{x}_0)\bm{K}^{-1}_i\bv_i,
\end{align*}
with $\bm{k}_i(\bm{x}_0)=[K(\bm{x}_0,\bm{x}_1;\btht_i),\dots,K(\bm{x}_0,\bm{x}_N;\btht_i)]^T$, and the scale parameter is
\begin{align*}
\hat{\sigma}_i^2(\bm{x}_0|\bv_i,\bm{\theta}_i)=\frac{(\beta_i+\psi_i)\Big(1-\bm{k}^T_i(\bm{x}_0)\bm{K}^{-1}_i\bm{k}_i(\bm{x}_0)\Big)}{\alpha_i+N}.
\end{align*}

Finally, the posterior distribution of $\sigma^2$ given
$\bY$ is 
\begin{align}\label{eq:posts2}
  \begin{aligned}
    \pi(\sigma^2|\bm{Y})&\propto \pi(\bY|\sigma^2)\pi(\sigma^2)\\
    &=
    (\sigma^2)^{-\frac{NL}{2}}\exp\left\{-\frac{\br^T\br}{2\sigma^2}\right\}(\sigma^2)^{-\frac{\alpha}{2}-1}\exp\left\{-\frac{\beta}{2\sigma^2}\right\}\\
    &=(\sigma^2)^{-\frac{NL}{2}-\frac{\alpha}{2}-1}\exp\left\{-\frac{\bm{r}^T\bm{r}+\beta}{2\sigma^2}\right\},
  \end{aligned}
\end{align}
where $\pi(\sigma^2)$ is the prior distribution of $\sigma^2$, and
%
$\bm{r}=\text{vec}(\bm{Y})-(\bI_N\otimes\bm{B})\text{vec}(\bV^{*T}),$
is the vectorization of residual matrix $\bY-\bB\bV^{*T}$.  The notation $\otimes$ represents the Kronecker product and the operator $\text{vec}(\cdot)$ performs vectorization for a matrix. Thus, $[\sigma^2|\bm{Y}]$ follows the inverse Gamma distribution $\text{IG}((NL+\alpha)/2,(\bm{r}^T\bm{r}+\beta)/2)$, and
\begin{align}\label{eq:int}
\hat{\sigma}^2=\underset{\sigma^2}{\mathrm{argmax}}~\pi(\sigma^2|\bm{Y})=\frac{1}{NL+\alpha+2}\left(\bm{r}^T\bm{r}+\beta\right).
\end{align}

The posterior predictive distribution of $\by(\bx_0)$ is given by
\begin{align*}
\pi\big(\bm{y}(\bm{x}_0)\big|\bV^*,\bm{\Theta},\sigma^2\big)\propto
\int_{\mathbb{R}^p}\pi\big(\bm{y}(\bm{x}_0)\big|\bm{c}(\bm{x}_0),\sigma^2\big)\prod_{i=1}^p\pi\big(c_i(\bm{x}_0)\big|\bv_i,\bm{\theta}_i\big)\prod_{i=1}^p
dc_i(\bm{x}_0),
\end{align*}
where $\bm{\Theta}=\{\bm{\theta}_1,\dots,\bm{\theta}_p\}$, and $\bm{c}(\bm{x}_0)= (c_1(\bm{x}_0), ..., c_p(\bm{x}_0))$. 
For a reasonably large value of $N$, a normal approximation can be imposed on the non-central $t_{N+\alpha_i}$ distribution of $[c_i(\bm{x}_0)|\bv_i,\bm{\theta}_i]$,  i.e.,
\begin{align}\label{eq:approx}
 \pi\big(c_i(\bm{x}_0)|\bv_i,\bm{\theta}_i\big)\approx
\mathcal{N}\left(\hat{c}_i(\bm{x}_0|\bv_i,\bm{\theta}_i),\hat{\sigma}_i^2(\bm{x}_0|\bv_i,\bm{\theta}_i)\right).
\end{align}
Furthermore, the results from Section 14.2 of \citet{gelman2014bayesian} can be summarized into Lemma~1 for further simplification of the predictive distribution of $\by(\bx_0)$.
\begin{lemma}
  Suppose $[\by|\bm{\beta},\sigma^2]\sim
  \mathcal{N}(\bX\bm{\beta},\sigma^2\bI_n)$ and $[\bm{\beta}]\sim
  \mathcal{N}(\bb,\bm{V})$, where $\by\in \mathbb{R}^n$,
  $\bm{\beta},\bb\in \mathbb{R}^m$, $\bX$ is an $n\times m$ matrix,
  and $\bm{V}$ is an $m\times m$ positive definite covariance matrix. Then,
  $[\by|\sigma^2]\sim\mathcal{N}(\bX\bb,\bX\bV\bX^T+\sigma^2\bI_n)$.
\end{lemma}

Combining (\ref{eq:md}) and (\ref{eq:approx}) with Lemma~1, we get
\begin{align}\label{eq:pred}
\pi(\bm{y}(\bm{x}_0)|\bV^*,\bm{\Theta},\sigma^2)\approx
\mathcal{N}\big(\bm{B}\hat{\bm{c}}(\bm{x}_0|\bV^*,\bm{\Theta}),\bB\bm{\Lambda}(\bV^*,\bm{\Theta})\bB^T+\sigma^2\bI_L\big),
\end{align}
where
$\hat{\bm{c}}(\bm{x}_0|\bV^*,\bm{\Theta})=[\hat{c}_1(\bm{x}_0|\bv_1,\bm{\theta}_1),\dots,\hat{c}_p(\bm{x}_0|\bv_p,\bm{\theta}_p)]^T,$ and
$\bm{\Lambda}(\bV^*,\bm{\Theta})= \text{diag}\big(\hat{\sigma}_1^2(\bm{x}_0|\bv_1,\bm{\theta}_1),\\ \dots,\hat{\sigma}_p^2(\bm{x}_0|\bv_p,\bm{\theta}_p)\big)$. The parameters $\bm{\Theta}$ and $\sigma^2$ cannot be
integrated out analytically, and \citet{kennedy2001bayesian} suggested using the MAP estimator into the predictive distribution. Following their paradigm, we plug $\hat{\sigma}^2$ and
\begin{align}\label{eq:theta}
  \hat{\bm{\theta}}_i=\underset{\bm{\theta}_i}{\mathrm{argmax}}~\pi(\bm{\theta}_i|\bv_i),~~~i=1,\dots,p,
\end{align}
into (\ref{eq:pred}) to obtain the approximate predictive
distribution
\begin{align}\label{eq:final}
\pi(\bm{y}(\bm{x}_0)|\bm{Y})
\approx \pi(\bm{y}(\bm{x}_0)|\bm{V}^*, \hat{\sigma}^2, \hat{\bm{\Theta}}) \approx
\mathcal{N}\big(\bm{B}\hat{\bm{c}}(\bm{x}_0|\bV^*,\hat{\bm{\Theta}}),\bB\bm{\Lambda}(\bV^*,\hat{\bm{\Theta}})\bB^T+\hat{\sigma}^2\bI_L\big).
\end{align}
where $\pi(\bm{\theta}_i|\bv_i)$ and $\pi(\sigma^2|\bm{Y})$
are given by (\ref{eq:post}) and (\ref{eq:posts2}), respectively. As a result, with the data $\bm{X}$ and $\bm{Y}$, the pre-specified hyperparameters $\balp$ and $\bbt$, as well as the threshold $\gamma$ in (\ref{eq:pp}), the SVD-based GP model fitting via Bayesian procedure provides the approximate predictive distribution in (\ref{eq:final}) with the plug-in estimates of $\hat{\sigma}^2$ in (\ref{eq:int}) and $\hat{\bm{\theta}}_i$'s in (\ref{eq:theta}). It shall be noted that,   the MAP estimator of $\bm{\theta}_i$'s can be shown to be robust \citep{gu2017robust}.  Algorithm 1 of the supplementary material summarizes the important steps in estimating the necessary parameters of the posterior predictive distribution (\ref{eq:final}) for a full SVD-based GP model.

Fitting the $i$th GP model ($1\le i\le p$) to $N$ training data points  involves numerous evaluations of the posterior (\ref{eq:post}), and the computation of $\bm{K}_i^{-1}$ and $|\bm{K}_i|$ requires $O(N^3)$ floating point operations (flops), which can quickly become infeasible even for moderately large $N$. Thus, we propose to use a localized SVD-based GP model that aims to achieve the same prediction accuracy at a substantially less computational cost.

\section{Local SVD-based GP Model}
\label{sec:loc-svdgp}

The main idea is to use a small subset of $n$ $ (\ll N)$ points instead of the entire training set of $N$ points for approximating the predicted response at an arbitrary $\bm{x}_0$ in the input space. Let $\bm{X}$ be the training set of $N$ points, and $\bX^{(n)}(\bm{x}_0)$ or $\bX^{(n)}$ (in short) denote the desired subset of $\bX$ which defines the $n$-point neighborhood of $\bm{x}_0$ contained in $\bX$. In this section, we discuss two methods of constructing this neighborhood set $\bX^{(n)}$.

The first one, called as the \emph{naive} approach, assumes the elements of the neighborhood set $\bX^{(n)}$ by finding $n$ nearest neighbors of $\bm{x}_0$ in $\bX$ as per the Euclidean distance in the \emph{k-nearest neighbor} method. The emulator obtained via fitting an SVD-based GP model (as described in Section~\ref{sec:svd-gpm}) to this local set of points is referred to as \emph{$k$-nearest neighbor SVD-based GP model} (in short, knnsvdGP). Though, knnsvdGP is computationally much cheaper than the \emph{full SVD-based GP model} (referred to as svdGP) trained on $N$ points, its prediction accuracy may not be satisfactory.

The second method (main focus of this paper) finds the neighborhood set $\bX^{(n)}(\bm{x}_0)$ (for every $\bm{x}_0$) using a greedy approach. \citet{gramacy2015local} developed a greedy
 sequential algorithm for constructing a neighborhood set for a scalar-valued simulator. In this paper, we propose a generalization of this algorithm for dynamic computer simulators. For every test point, the generalized greedy algorithm finds a local set of points in the training set to build an SVD-based GP model. Such model is referred to as the {\em local approximate SVD-based GP model} (in short, lasvdGP).

\subsection{Local Approximate SVD-based GP Model}
\label{sec:gNN}

For every given  $\bm{x}_0$ in the input space, the proposed approach starts with finding a smaller neighborhood set $\bX^{(n_0)}(\bm{x}_0)$, which consists of $n_0$ $(<n)$ nearest neighbors of $\bm{x}_0$ in $\bX$ (with respect to the Euclidean distance). This step is the same as in knnsvdGP with $n$ replaced by $n_0$. The remaining $n-n_0$ neighborhood points are chosen sequentially one at-a-time by optimizing a merit-based criterion over the input space. The prime objective is to reduce the overall prediction error. The key steps of the proposed lasvdGP approach is summarized in Algorithm 2 of the supplementary material.

Let $k$ denote the current number of points in the neighborhood set, $\bX^{(k)}$ and $\bm{X} \backslash \bX^{(k)}$ be the sets of selected and unselected (remaining) training points, respectively, and $\hat{\bm{\Theta}}^{(k)}=\{\hat{\bm{\theta}}_1^{(k)},\dots,\hat{\bm{\theta}}^{(k)}_p\}$ be the estimated correlation parameters using $\bX^{(k)}$ and $\bY(\bX^{(k)})$.  Then the next follow-up point in the neighborhood set is chosen as
\begin{align*}
  \bm{x}^*_{k+1}=\underset{\bm{x}\in
  \bm{X}\backslash\bX^{(k)}}{\mathrm{argmin}}\:J(\bx_0,\bx),
\end{align*}
\noindent where \begin{align}\label{eq:l2}
  \resizebox{\hsize}{!}{%
  $J(\bx_0,\bx)=\tE\bigg\{\tE\Big[\big\|\by(\bx_0)-\hat{\by}(\bx_0|
\bc(\bx),\bV^{*(k)},\hbT^{(k)})\big\|^2\Big|\bc(\bx),\bV^{*(k)},\hbT^{(k)},(\hat{\sigma}^{(k)})^2\Big]\bigg|\bV^{*(k)},\hbT^{(k)},(\hat{\sigma}^{(k)})^2\bigg\},$%
}
\end{align}
with
\begin{align}\label{eq:exp-yx0}
  \begin{aligned}
  \hat{\bm{y}}\big(\bm{x}_0|
  \bc(\bx),\bV^{*(k)},\hat{\bm{\Theta}}^{(k)}\big)&=\text{E}\left[\bm{y}(\bm{x}_0)\middle|\bc(\bx),\bV^{*(k)},\hat{\bm{\Theta}}^{(k)},(\hat{\sigma}^{(k)})^2\right]\\
  &=\bB^{(k)}\hbc\big(\bx_0|\bc(\bx),\bV^{*(k)},\hbT^{(k)}\big),
  \end{aligned}
\end{align}
where $\bB^{(k)}$ and $\bV^{*(k)}$ are the matrices of basis vectors and the right singular vectors, $p_k$ is the number of bases selected in this iteration, and $\bc(\bx)=[c_1(\bx),\dots,c_{p_k}(\bx)]^T$ with $c_i \sim \text{GP}(0,\sigma_i^2 K(\cdot,\cdot;\bm{\theta}_i^{(k)}))$. The predictive mean vector of coefficients $\hbc(\bx_0|\bc(\bx),\bV^{*(k)},\hbT^{(k)})$ is calculated in the exact same way as (\ref{eq:pred}) except
$[(\bV^{*(k)})^T,\bc(\bx)]^T$ and $\bTht^{(k)}$ are used in place of $\bV^*$ and $\bTht$, respectively.

This $J$-criterion is a generalization of the active learning Cohn (ALC) criterion (\citet{cohn1996active};\citet{cohn1996neural};\citet{gramacy2015local})
\begin{align*}
  \int_{\bx}\Big[\int_y\big(\hat{y}(\bx)-y(\bx)\big)^2dP(y|\bx)\Big] dP(\bx),
\end{align*}
where $y(\bx)$ and $\hat{y}(\bx)$ are the observed and predicted scalar-valued outputs, respectively, at input $\bx$, $P(y|\bx)$ is the approximate predictive distribution,
and the marginal distribution $P(\bx)$ is uniform. For dynamic computer simulators, we use $L_2$ norm discrepancy
instead of the squared error.

The closed form expression for $J(\bx_0,\bx)$ can be derived by taking the outer expectation in (\ref{eq:l2}) with respect to the approximate posterior distribution in (\ref{eq:pred}) and substituting ($\bV^{*(k)}$, $\hbT^{(k)}$, $(\hat{\sigma}^{(k)})^2$) for ($\bV^*$, $\bTht$, $\sigma^2$). Similarly, the expectation in (\ref{eq:exp-yx0}) is computed with respect to (\ref{eq:pred}), and by substituting ($[(\bV^{*(k)})^T,\bc(\bx)]^T$, $\hbT^{(k)}$, $(\hat{\sigma}^{(k)})^2$) for ($\bV^*$, $\bTht$, $\sigma^2$). Proposition \ref{propjl} states the closed form expression of $J(\bx_0,\bx)$, and the proof is shown in the Appendix~A of the supplementary materials.\\

\begin{proposition}\label{propjl}
Suppose the expectations in (\ref{eq:l2}) and (\ref{eq:exp-yx0}) are taken with respect to the approximate predictive distribution (\ref{eq:pred}). Then, for any $\bx\in\bX\backslash\bX^{(k)}$
\begin{align*}
  J(\bx_0,\bx) = (\hat{\sigma}^{(k)})^2L+\sum_{i=1}^{p_k}(d_i^{(k)})^2\hat{\sigma}^2_i\big(\bm{x}_0|\bx,\bv^{(k)}_i,\hat{\bm{\theta}}_i^{(k)}\big),
\end{align*}
where $d_i^{(k)}$ is the $i$th  largest singular value of
$\bY^{(k)}$,
\begin{align*}
  &\hat{\sigma}^2_i\big(\bm{x}_0|\bx,\bv^{(k)}_i,\hat{\bm{\theta}}_i^{(k)}\big)=
    \frac{\rho_i^{(k)}(\bx_0,\bx)}{\alpha_i+k}\big(\beta_i+\frac{\alpha_i+k}{\alpha_i+k-1}\psi_i^{(k)}\big),\\
  &\rho^{(k)}_i(\bx_0,\bx)=1-\tbk_i(\bx_0,\bx)^T\tbK_i^{-1}(\bx)\tbk_i(\bx_0,\bx),\nonumber\\
  &\psi^{(k)}_i=(\bv^{(k)}_i)^T(\bK_i^{(k)})^{-1}\bv^{(k)}_i,\nonumber\\
&\tbk_i(\bx_0,\bx)=[K(\bx_0,\bx^{(k)}_1;\hbt^{(k)}_i),\dots,K(\bx_0,\bx^{(k)}_k;\hbt^{(k)}_i),K(\bx_0,\bx;\hbt^{(k)}_i)]^T,\nonumber\\
&\tbK_i(\bx)=\left[
  \begin{array}{cc}
    \bK^{(k)}_i&\bk^{(k)}_i(\bx)\\
    \bk^{(k)}_i(\bx)^T& 1 \\
  \end{array}
  \right],\nonumber
\end{align*}
with $\bv_i^{(k)}$ being the $i$th column of $\bV^{*(k)}$, for
$i=1,\dots,p_k$, $\bx_j^{(k)}$ being the $j$th point of
$\bX^{(k)}$ for $j=1,\dots,k$, $\bK^{(k)}_i$ being a
$k\times k$ matrix with $K(\bx_j^{(k)},\bx_l^{(k)};\hbt^{(k)}_i)$, as the $(j,l)$th entry, and $\bk_i^{(k)}(\bx)=[K(\bx,\bx_1^{(k)};\hbt^{(k)}_i),\dots,K(\bx,\bx_k^{(k)};\hbt^{(k)}_i)]^T$.\\
\end{proposition}

As $(\hat{\sigma}^{(k)})^2L$ is a constant with respect to
$\bx$, finding $\bm{x}^*_{k+1}$, by minimizing the $J$-criterion in Proposition~\ref{propjl}, is equivalent to obtaining
\begin{align}\label{eq:wsum}
  \bm{x}^*_{k+1}=\underset{\bm{x}\in
  \bm{X}\backslash\bX^{(k)}}{\mathrm{argmin}}\Big[\sum_{i=1}^{p_k}(d^{(k)}_i)^2\hat{\sigma}^2_i\big(\bm{x}_0|\bx,\bv^{(k)}_i,\hat{\bm{\theta}}_i^{(k)}\big)\Big].
\end{align}

Note that the simplified design criterion in (\ref{eq:wsum}) turns out to be the weighted sum of the predictive variance of the singular vector coefficients, where the weights are $(d^{(k)}_i)^2$ which represents the total variation explained by the $i$th singular vector basis. Therefore, the chosen follow-up point $\bm{x}^*_{k+1}$ minimizes the expected $L_2$ prediction error at $\bx_0$ evaluated at stage $k$.

As compared to knnsvdGP, the proposed algorithm, lasvdGP, requires many more GP model fitting steps, which increase the computational cost, however, it is still substantially faster than the svdGP implementation. The matrix inverse updating procedure employed in \citet{hager1989updating} (also used in \citet{gramacy2015local}) can be used to achieve further time saving from $O(k^3)$ to $O(k^2)$  in
  evaluating $J(\bx_0,\bx)$ of Proposition~1 for each
  $\bx\in\bX\backslash \bX^{(k)}$, where $k=n_0,\dots,n-1$, is the
  number of neighborhood points in the current neighborhood set
  $\bX^{(k)}$. This is because, the evaluation of the $J$-criterion requires inverting $(k+1)\times (k+1)$ correlation matrices
  $\{\tilde{\bm{K}}_i(\bx)\}_{i=1}^p$, and applying the matrix inverse update, we have
\begin{align*}
    \tilde{\bm{K}}_i(\bx)^{-1}=\left[
    \begin{array}{cc}
      (\bm{K}_i^{(k)})^{-1}+\bm{g}_i\bm{g}_i^T\phi_i&\bm{g}_i\\
      \bm{g}_i^T& \phi_i^{-1}
    \end{array}
\right],
  \end{align*}
  where $\bm{g}_i=-(\bm{K}_i^{(k)})^{-1}\bm{k}_i^{(k)}(\bx)/\phi_i$ and
  $\phi_i=1-\bm{k}_i^{(k)}(\bx)^T(\bm{K}_i^{(k)})^{-1}\bm{k}_i^{(k)}(\bx)$.
  Thus, the computation  of  $ \tilde{\bm{K}}_i(\bx)^{-1}$ attributes to computing both $(\bm{K}_i^{(k)})^{-1}$ and $\bm{g}_i$, the former of which has been  calculated and stored in the process of estimating range parameters and thus no additional computing time is required for evaluating $(\bm{K}_i^{(k)})^{-1}$ in the $J$-criterion. On the other hand, the evaluation of $\bm{k}_i^{(k)}(\bx)$ in $\bm{g}_i$ requires $O(k)$ time, and the complexity of the matrix multiplication $(\bm{K}_i^{(k)})^{-1} \bm{k}_i^{(k)}(\bx)$ is $O(k^2)$. Note that the anticipated boost in the prediction accuracy at the cost of a small increase in the computational cost is perhaps worth it. The computational complexities of the two methods  knnsvdGP and lasvdGP are more extensively discussed in Section~\ref{sec:cc}.

\subsection{Computational Complexity}
\label{sec:cc}

In this section, we discuss  the computational complexity of (1) full SVD-based GP model (svdGP), (2) $k$-nearest neighbor  SVD-based GP model (knnsvdGP), and (3)  local approximate SVD-based GP model (lasvdGP). For this comparison, let $\bm{X}$ contain $N$ training points, $\bm{X}^*$ consist of $M$ test points, $L$ be the length of time series response, each neighborhood set in knnsvdGP and lasvdGP consists of $n$ training points, and $N>L>n$ (assuming $N$ is large). Furthermore, we only compute the diagonal entries of the predictive covariance matrix of $\by(\bx_0)$, i.e., the marginal predictive variances, for each $\bm{x}_0\in\bm{X}^*$.

(1) \textbf{svdGP:} A single call of the empirical Bayesian inference for SVD-based GP model on the full training data requires $O(N^3)$ floating point operations (flops). Since we assume $N>L$, the estimation of $\bTht$ is the dominant part of the empirical Bayesian inference computation, which requires $O(N^3)$ flops. The complexity of the prediction step is $O(M(N^2+L))=O(MN^2)$, and thus, the total cost of svdGP is $O(N^2\max\{M,N\})$.

(2) \textbf{knnsvdGP:} For each $\bx_0\in \bX^*$, the neighborhood set construction needs $O(nN)$ flops, and one call of singular value decomposition takes
$O(nL\min\{n,L\})$ flops (\citet{gentle2007matrix}), which is
$O(n^2L)$ since $n<L$ is assumed. The estimation of $\hat{\sigma}^2$ and $\hbT$ based on $n$ neighborhood points requires $O(nL)$ and $O(n^3)$ flops, respectively. That is, the cost of the empirical Bayesian inference for SVD-based GP models based on $n$ neighborhood points is $O(n^2L+nL+n^3)=O(n^2L)$. Furthermore, the computational complexity of the prediction step is $O(n^2+L)$. Consequently, the total cost of knnsvdGP algorithm is $O(Mn\max\{nL,N\})$.

(3) \textbf{lasvdGP:}  The cost of empirical Bayesian inference for SVD-based GP models based on $k$ neighborhood points is $O(k^2L)$, as in knnsvdGP, $n_0 \leq k \leq n$. Optimization of the $J$-criterion costs $O(k^2N)$ (as per the quick update formula by \citet{gramacy2015local}). Thus the cost of building a  local approximate SVD-based GP model at the $k$-th iteration is $O(k^2N)$, $k=n_0,...,n-1$, and thus the entire process of fitting  a local approximate SVD-based GP model requires $\sum_{k=n_0}^{n-1}O(k^2N)=O(n^3N)$ flops. Note that the prediction cost in
  lasvdGP is not significant compared to the neighborhood selection
  and inference of the GP models. As a result, the total cost of this algorithm is $O(n^3NM)$.

Table \ref{tab:cost} summarizes the computational complexity of the three methods.


\begin{table}[h!]
  \centering
  \caption{The computational cost of fitting GP models under the three methods.}
  \begin{tabular}{lccc}
\hline
    Method  & svdGP & knnsvdGP                & lasvdGP      \\
\hline
    Cost   & $O(N^2\max\{M,N\})$ & $O(Mn\max\{nL,N\})$ & $O(n^3NM)$  \\
\hline
  \end{tabular}
\label{tab:cost}
\end{table}

It is easy to see that lasvdGP is computationally more expensive than knnsvdGP, however, the gain in the prediction accuracy is perhaps worth more. Assuming $M=O(N)$, it is also straightforward to notice that knnsvdGP and lasvdGP are substantially faster than  svdGP (full model) as long as $n=O(N^{1/3})$.

\subsection{Implementation}
\label{sec:imp}

Both local SVD-based GP models (knnsvdGP and
  lasvdGP) are run in a parallel computing environment using the \proglang{R} package
   \pkg{parallel} (\citet{R2015}).
One quick option is to divide the job into $M$ parts and fit independent local GP models. In contrast, svdGP models cannot be parallelized in such an easy manner, except the prediction component. Of course, one could use parallelization for SVD of $\bY$, and/or computing the determinant and inverse of the correlation matrices within the optimization step.

We implemented the three methods in \proglang{R} (\citet{R2015}). The
parallelization of the empirical Bayesian estimation and the prediction at $M$ untried inputs are implemented via the package {\em parallel}. The optimization in empirical Bayesian inference for all the three methods is performed with the assistance of the \pkg{laGP} package with default priors (\citet{gramacy2015lagp}). We shall also mention that in searching for the best follow-up point $\bm{x}^*_{k+1}$ from the candidate set $\bm{X} \backslash \bX^{(k)}$,  we adopt the limit search scheme suggested by (\citet{gramacy2015lagp}) instead of the exhaustive search. This allows us to save tremendous computational time without sacrificing prediction accuracy, as indicated by our empirical studies.

Fitting GP models to a large number of observations in low input dimension can often run into numerical instability due to near-singularity, and typically a small nugget is used in the correlation structure to address this numerical issue (e.g., \citet{ranjan2011computationally}; \citet{gramacy2012cases}; \citet{peng2014choice}). Similar to \cite{gramacy2015local}, we fix the nugget $\eta$ at a pre-determined small value to avoid near-singularity issue of the  correlation matrix frequently emerged in the Gaussian correlation family \citep{gu2017robust}.

\section{Applications}
\label{sec:simu-local}

In this section, we consider two examples with different test functions that represent dynamic computer models. We also consider a real-life application where the computer simulator (TDB model) generates population growth curve.  The complexity of the examples considered here range from $N=10,000$ to $30,000$ (size of the training set), and $q=3$ to $11$ (input dimension).

The performance of the three methods svdGP, knnsvdGP and lasvdGP is evaluated by comparing the normalized mean squared prediction error (NMSPE),
\begin{align}\label{eq:mspe}
  \text{NMSPE}(\bm{x})=\frac{\sum_{t=1}^L\big(y_t(\bm{x})-\hat{y}_t(\bm{x})\big)^2}{
  \sum_{t=1}^L\big(y_t(\bm{x})-\bar{y}(\bm{x})\big)^2},
\end{align}
and the proper scoring rule (\citet{gneiting2007strictly}) defined as
    \begin{align}\label{eq:prop-score}
      S(P_{\hat{\by}(\bx)},\by(\bx))=
      -\frac{1}{L}\sum_{t=1}^L\frac{(y_t(\bx)-\hat{y}_t(\bx))^2}{\hat{\sigma}^2_t(\bx)}-\frac{1}{L}\sum_{t=1}^L\log\hat{\sigma}_t^2(\bx),
    \end{align}
where $P_{\hat{\by}(\bx)}$ is the predictive distribution of the response at $\bx$, $\by(\bx)=[y_1(\bx),\dots,y_L(\bx)]^T$ is the (typically unknown) true response time-series at $\bx$,  $\hat{y}_t(\bm{x})$ is the corresponding predicted mean response, and $\hat{\sigma}_t^2(\bx)$ is the associated variance given by the $t$th diagonal  entry of $\bm{B}\bm{\Lambda}(\bm{V}^*,\hat{\bm{\Theta}})\bm{B}^T+\hat{\sigma}^2\bm{I}_L$. Furthermore, the temporal mean is given by $\bar{y}(\bm{x})=\sum_{t=1}^Ly_t(\bx)/L$. As model ranking criteria, the objective is to minimize average NMSPE and maximize the mean proper scoring rule.

For all these methods, we use the default priors of the \proglang{R} package \pkg{laGP}, i.e., the vague scale-invariant priors (\citet{gramacy2005thesis}) with $\alpha_i$'s, $\beta_i$'s, $\alpha$ and $\beta$ set to be 0, and for the correlation parameters $\bm{\theta}_i$'s, the  priors are explained at the beginning of Section 2.2.
We adopt zero-mean function in all GP models, apply the models to the normalized outputs that have zero mean and add the mean back for prediction.
For the simulated test functions, Examples~1 and 2, we repeat the emulation procedure 50 times with different (randomly chosen) training and test data sets and compare the average performance. For the real application in Example~3, we used Monte Carlo cross-validation approach for quantifying uncertainty in the prediction process (\citet{shao1993linear}).

\subsection{Example 1 (\citet{forrester2008})}
\label{sec:exp-forretal}

Consider the following test function with 3-dimensional inputs to generate simulator responses with time-series outputs,
\begin{align}\label{eq:ex2}
f(\bm{x},t) = (x_1t-2)^2\sin(x_2t-x_3),
\end{align}
where $\bm{x}=(x_1,x_2,x_3)^T\in[4,10]\times[4,20]\times[1,7]$, and
$t\in[1,2]$ is on a 200-point equidistant time-grid.

\begin{figure}[h!]\centering
	\includegraphics[width=6.0in]{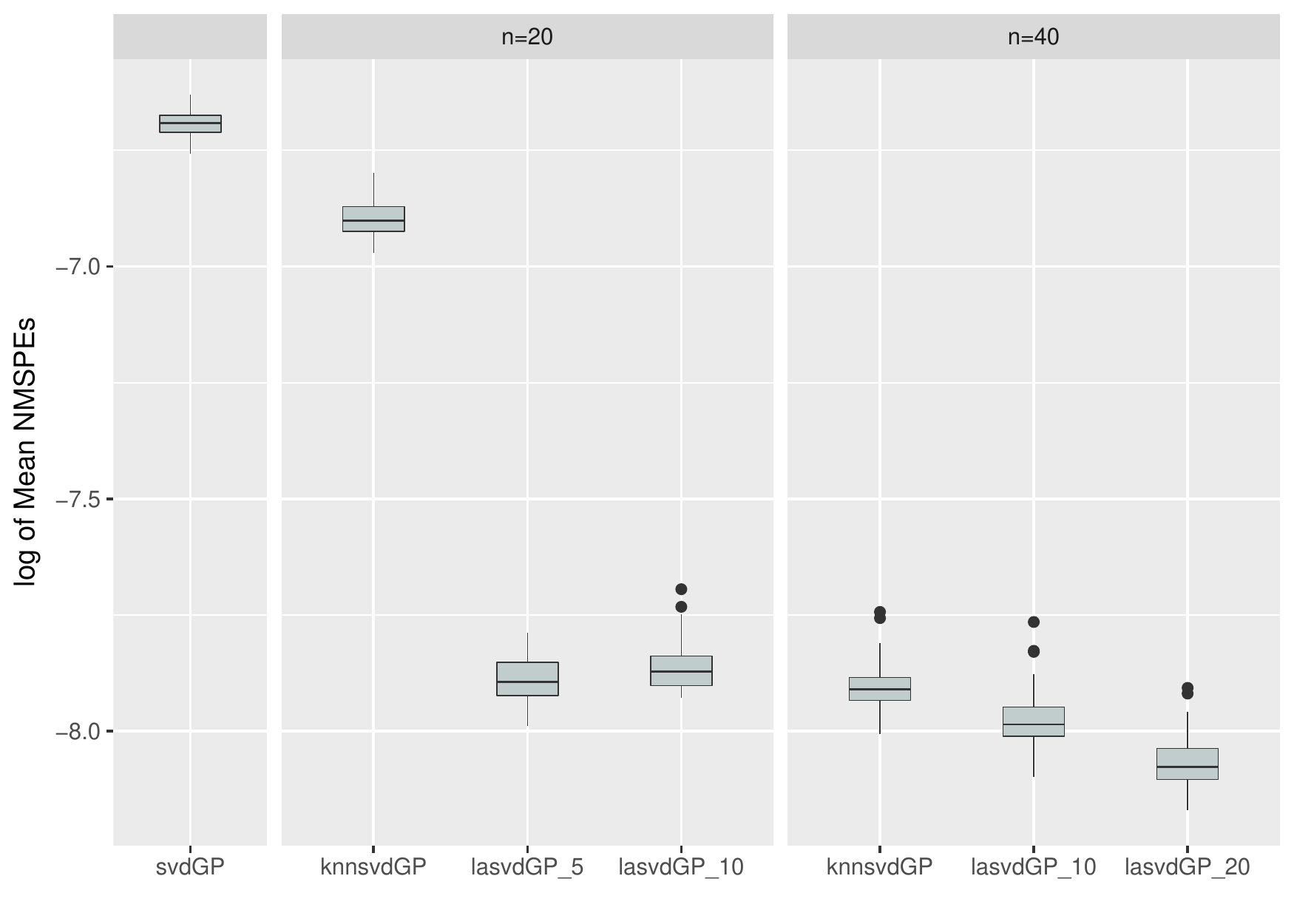}
	\caption{The boxplots of the log of mean NMSPE computed from 2,000 test points over 50 simulations for the computer simulator (\ref{eq:ex2}). The proposed lasvdGP approach achieves much smaller log of mean NMSPE values than the competitors.} \label{fig:ex2}
\end{figure}

\begin{figure}[h!]\centering
	\includegraphics[width=6.0in]{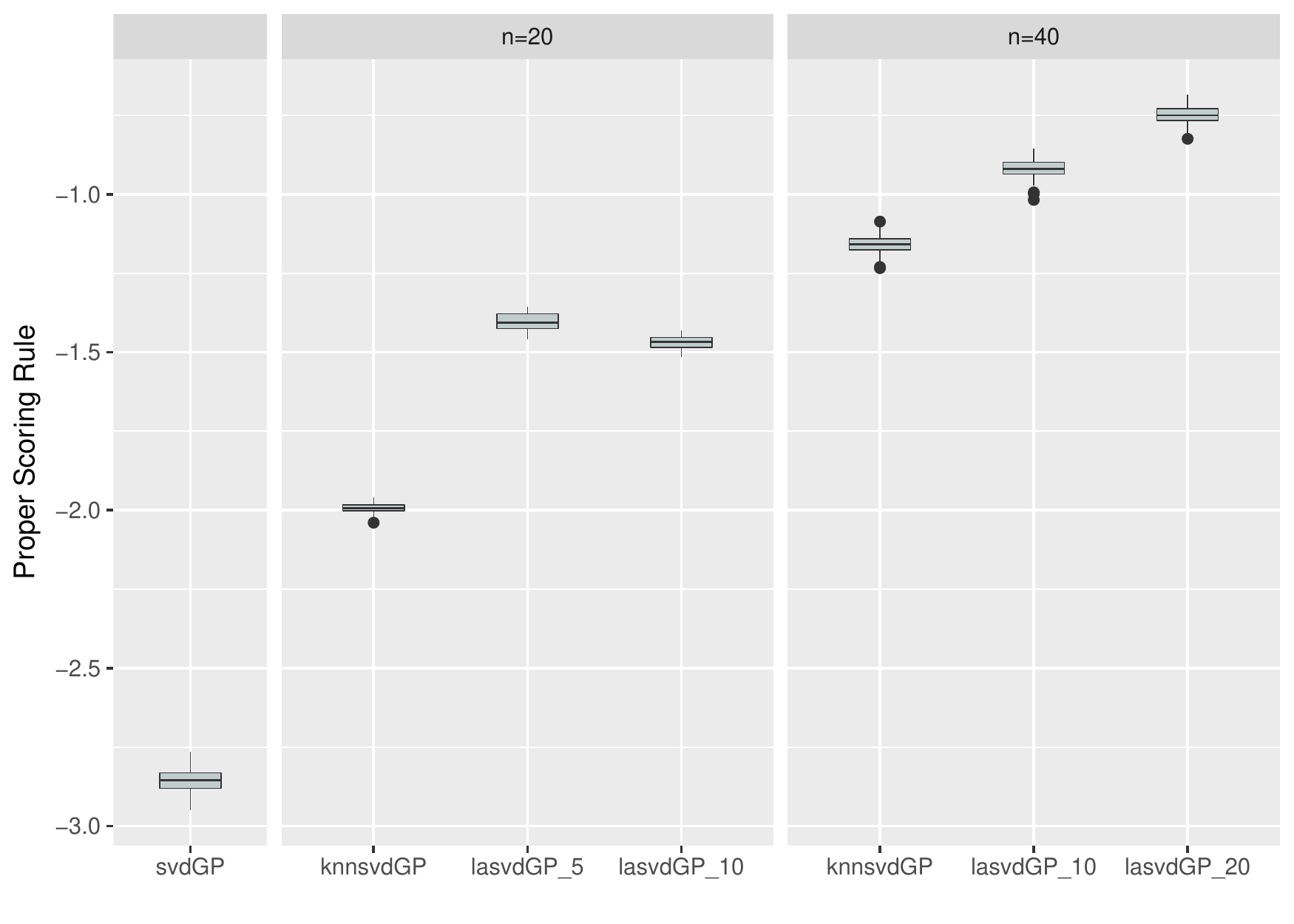}
	\caption{The boxplots of the mean proper scoring rule computed from 2,000 test points over 50 simulations for the computer simulator (\ref{eq:ex2}). The proposed lasvdGP approach achieves higher values of mean proper scoring rule  than the competitors. } \label{fig:ex2_score}
\end{figure}

For each of 50 replications, we randomly generate the training data of size 10,000 and the test data of size 2,000 using random Latin hypercube designs (LHDs) (\citet{mckay1979comparison}) from the input space $[4,10]\times[4,20]\times[1,7]$. The local approximate methods are implemented on the neighborhood sets of size $n=20$ and $40$ points. For lasvdGP, we assume the initial neighborhood size to be $n_0=\ceil{n/4}$, and $\ceil{n/2}$, where $\ceil{x}$ represents the smallest integer greater than or equal to $x$. Figures~\ref{fig:ex2} and \ref{fig:ex2_score} summarize the log of mean NMSPE and mean proper scoring rule values, respectively, for different models. Notation: lasvdGP$\_{n_0}$ denotes that the proposed method uses $n_0$ points in the initial neighborhood set chosen as nearest points based on Euclidean distance, and the remaining $n-n_0$ points are chosen sequentially by optimising the $J$-criterion.


Figures~\ref{fig:ex2} and \ref{fig:ex2_score} reveal that the proposed algorithm outperforms its naive counterpart irrespective of the total neighborhood size $(n)$. As the neighborhood size gets large, the prediction accuracy of the local approximation algorithms improves.

For a fixed data set of size 10,000, we also computed Monte Carlo cross-validation based values for the two measures, log of mean NMSPE and mean proper scoring rule, and compared the three models. We considered one-fifth of the data as the test set and the remaining as the training set. The  boxplots of the two measures over 50 random splits of the data show the similar trend as in Figures~\ref{fig:ex2} and \ref{fig:ex2_score}. 

\subsection{Example 2 (\citet{bliznyuk2008bayesian})}
\label{sec:exp-largeenv}

Consider the environmental model in
\citet{bliznyuk2008bayesian} which models a pollutant spill caused by a chemical accident. The simulator output is given by
\begin{align}\label{eq:ex3}
  \begin{aligned}
    &f(\bm{x},t)= \frac{M}{\sqrt{
        Dt}}\exp\left(\frac{-s^2}{4Dt}\right)+\frac{M}{\sqrt{
        D(t-\tau)}}\exp\left(-\frac{(s-L)^2}{4D(t-\tau)}\right)I(\tau<t),
  \end{aligned}
\end{align}
where $\bm{x}=(M,D,L,\tau,s)^T$, $M$ denotes the mass of pollutant spilled at each location, $D$ is diffusion rate in the channel, $L$ is location of the second spill, $\tau$ is time of the second spill, $\bm{x}\in [7,13]\times[0.02,0.12]\times[0.01,3]\times[30.01,30.295]\times[0,3]$, and $t\in [0.3,60]$ is on a regular 200-point equidistant time grid.

\begin{figure}[h!]\centering
  \includegraphics[width=6in]{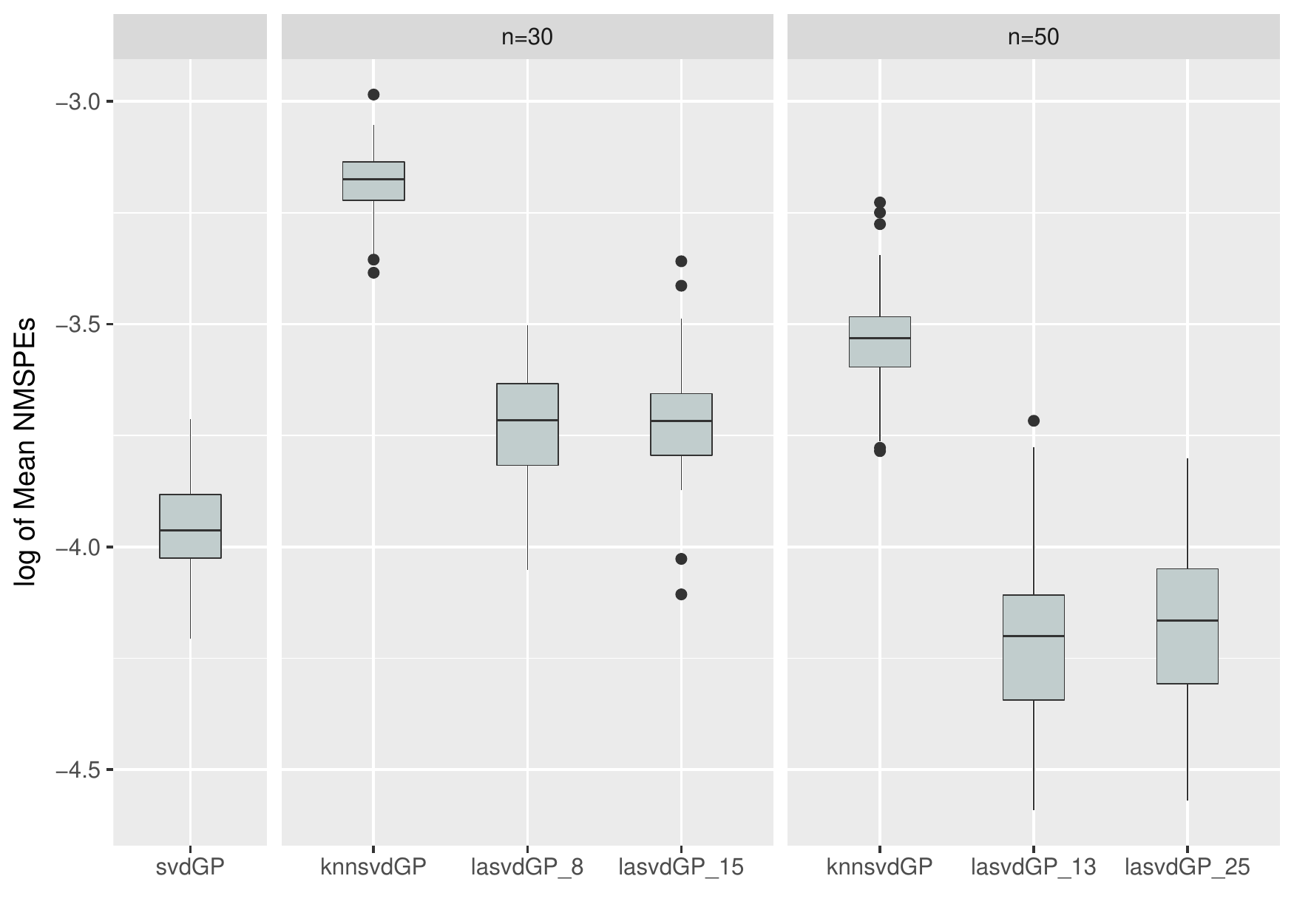}
\caption{The boxplots of the log of mean NMSPEs computed from 2,000 test points over 50 simulations for the simulator given by (\ref{eq:ex3}). The proposed lasvdGP approach achieves much smaller log of mean NMSPE values than the competitors. }\label{fig:ex4}
\end{figure}

\begin{figure}[h!]\centering
  \includegraphics[width=6.0in]{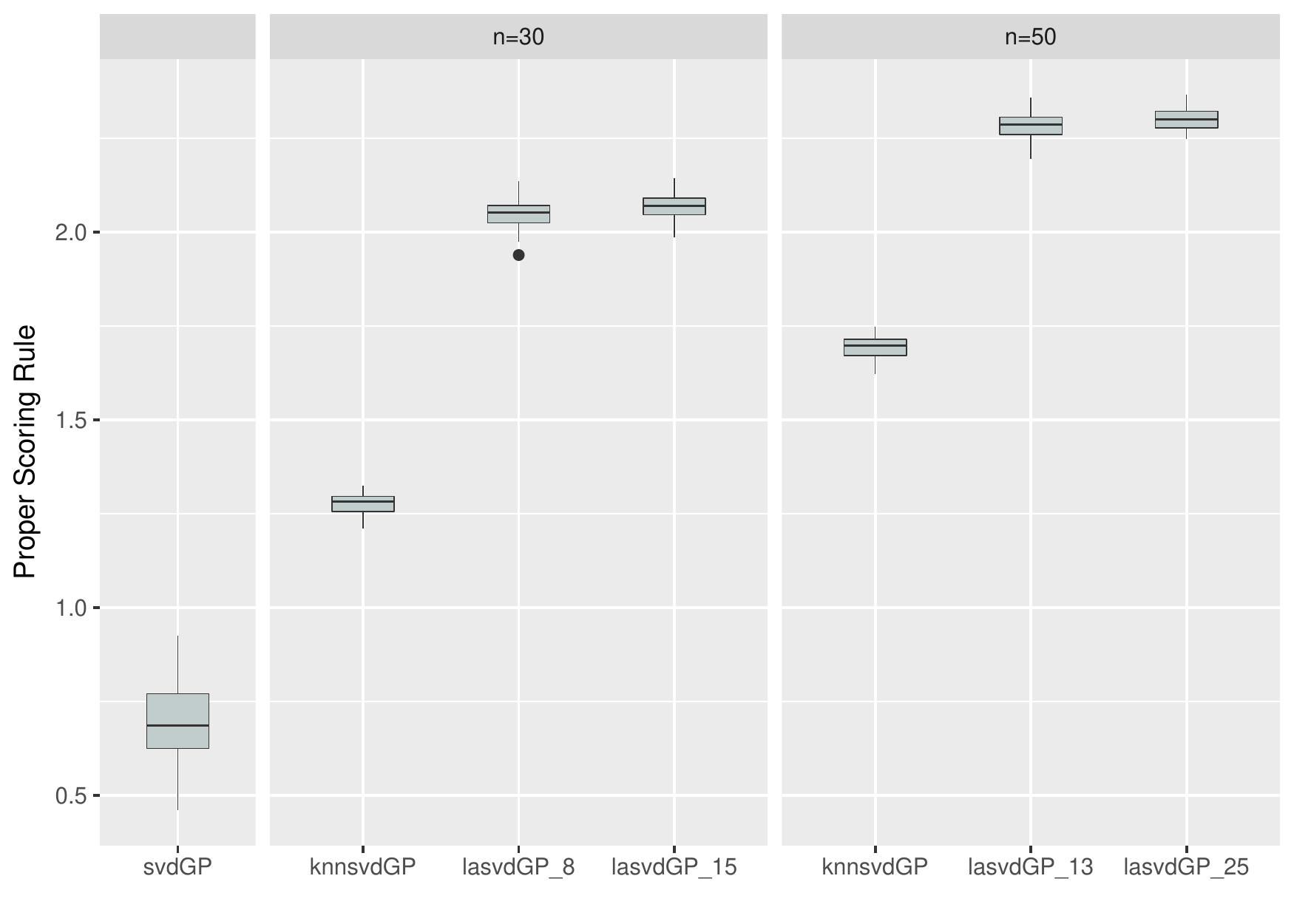}
\caption{The boxplots of mean proper scoring rule computed from 2,000 test points over 50 simulations for the simulator given by (\ref{eq:ex3}). The proposed lasvdGP approach achieves higher values of mean proper scoring rule  than the competitors.}\label{fig:ex4_score}
\end{figure}

In this example as well, we use the training data of size $N=10,000$ and the test data of size $M=2,000$ obtained using a random LHD. Similar to the previous example, Figures~\ref{fig:ex4} and \ref{fig:ex4_score} display the boxplots of 50 log of mean NMSPEs and mean proper scoring rule values, respectively, computed over the test set.

\begin{figure}[h!]\centering
  \includegraphics[width=5.0in]{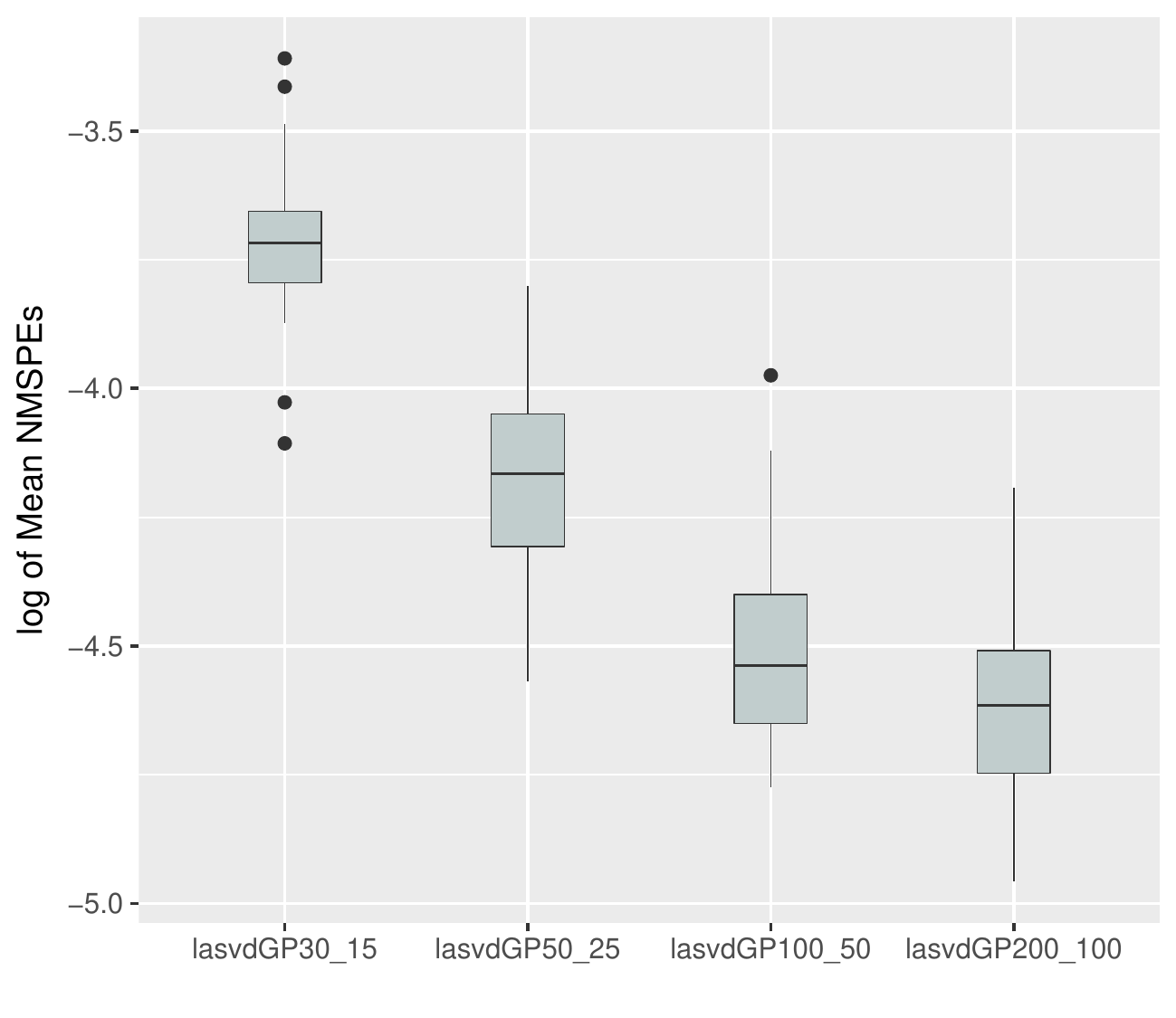}
\caption{The boxplots of the log of mean NMSPEs computed using 2,000 test points over 50 simulations with $n=30,50,100,200$ and $n_0 = n/2$ for model (\ref{eq:ex3}). As the neighborhood size increases, the log of mean NMSPE values by the proposed lasvdGP approach decrease, resulting in more accurate prediction.}\label{fig:ex4_box}
\end{figure}

Focussing on the local GP models, Figures~\ref{fig:ex4} and \ref{fig:ex4_score} demonstrate that the proposed approach (lasvdGP) is more accurate than the naive one (knnsvdGP), and $n=50$ exhibits more accurate prediction than $n=30$. As in the previous example, the Monte Carlo cross-validation approach shows consistent findings. To investigate this further, we compared the prediction accuracy of lasvdGP for different $n$ and $n_0=\lceil n/2\rceil$, Figure~\ref{fig:ex4_box} summarizes the findings.

Figure~\ref{fig:ex4_box} shows the expected increasing trend of the average prediction accuracy. Though the prediction accuracy increases with $n$, the rate of increment in the accuracy slows down as $n$ increases, and more importantly, note that fitting a lasvdGP model, requires $O(n^3NM)$ flops, which becomes prohibitively large very quickly.


\subsection{Example 3 (TDB simulator - \citet{teismann2009})}
\label{sec:real-data}

The two-delay blowfly (TDB) model (\citet{teismann2009})
simulates European red mites (ERM) population dynamics under predator-prey interactions in apple orchards via numerically solving the Nicholson's blowfly differential equation (\citet{gurney1980}). Unmanaged ERM population growth could incur massive infestation which inflicts heavy loss in apple industry. Therefore, the monitoring and subsequent intervention of ERM population dynamics is of vital importance for apple orchards management. The objective here is to emulate this simulator for deeper insight in  the process.

\begin{figure}[h!]
  \centering
  \includegraphics[width=6.5in]{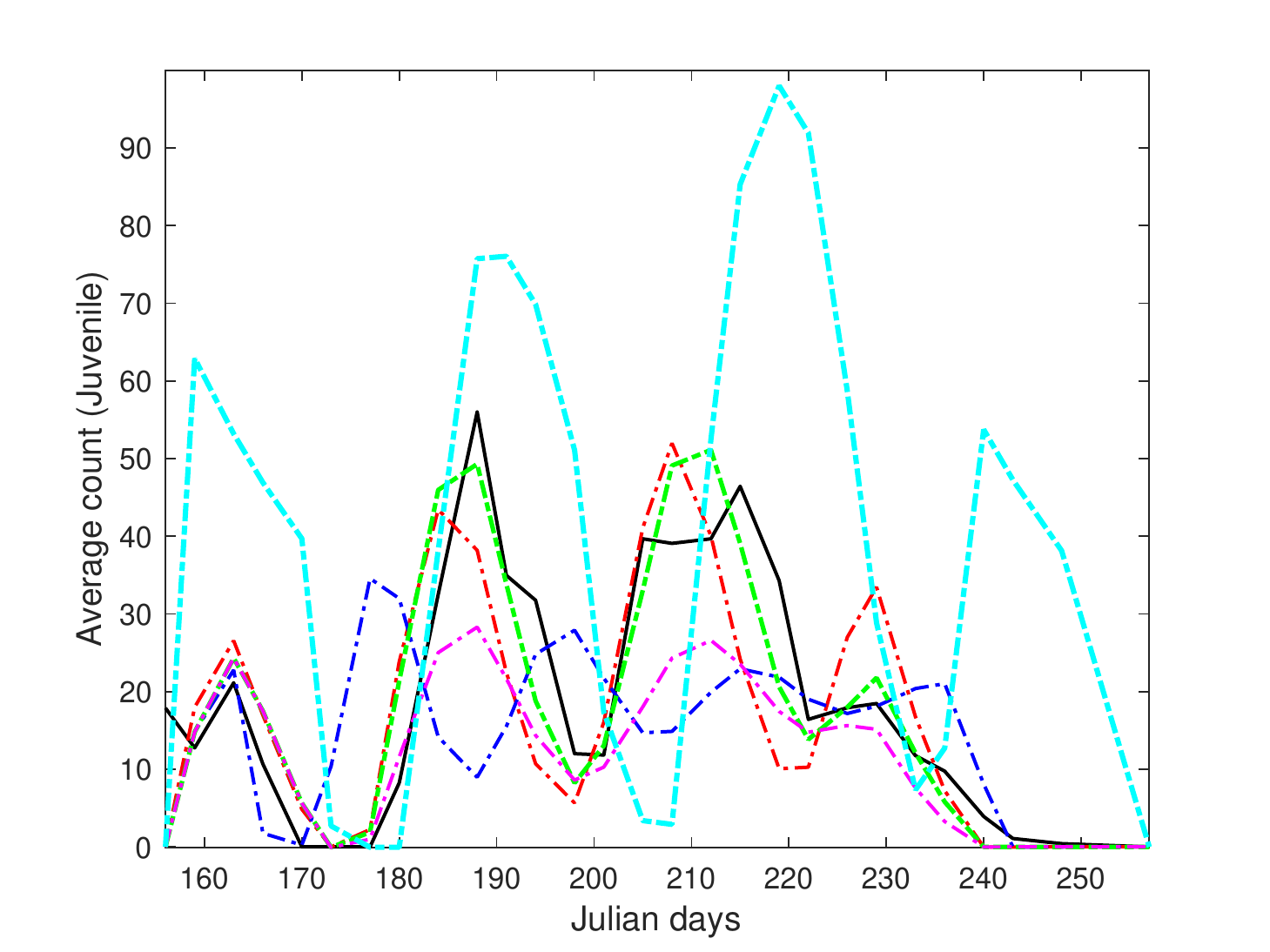}
  \caption{Juvenile ERM population dynamics as outputs of the TDB model at five different inputs. The solid curve shows the field data, and the dashed curves show the TDB outputs.}
  \label{fig:tdb-show}
\end{figure}

The TDB model takes eleven input variables (e.g., death rates for different stages, fecundity, hatching time, survival rates, and so on) and returns the time series (at 28 time points) of ERM population evolutions at three stages, i.e., eggs, juveniles and adults (see \citet{ranjan2016inverse} for details). In this paper, we focus on the population dynamics of juveniles. Figure \ref{fig:tdb-show} shows the model output at five randomly chosen input points.
%

%

\begin{figure}[h!]
  \centering
  \includegraphics[width=3.15in]{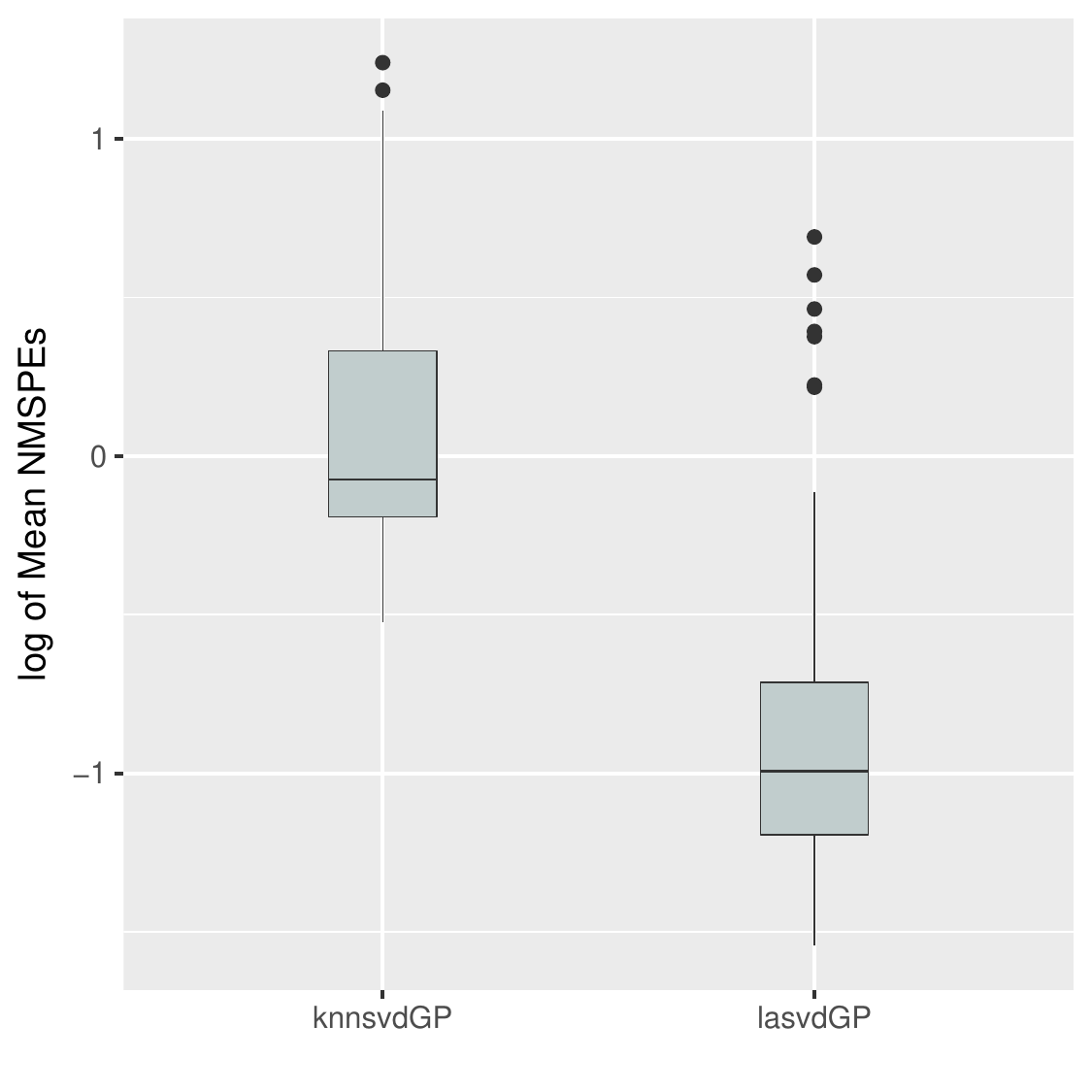}
  \includegraphics[width=3.15in]{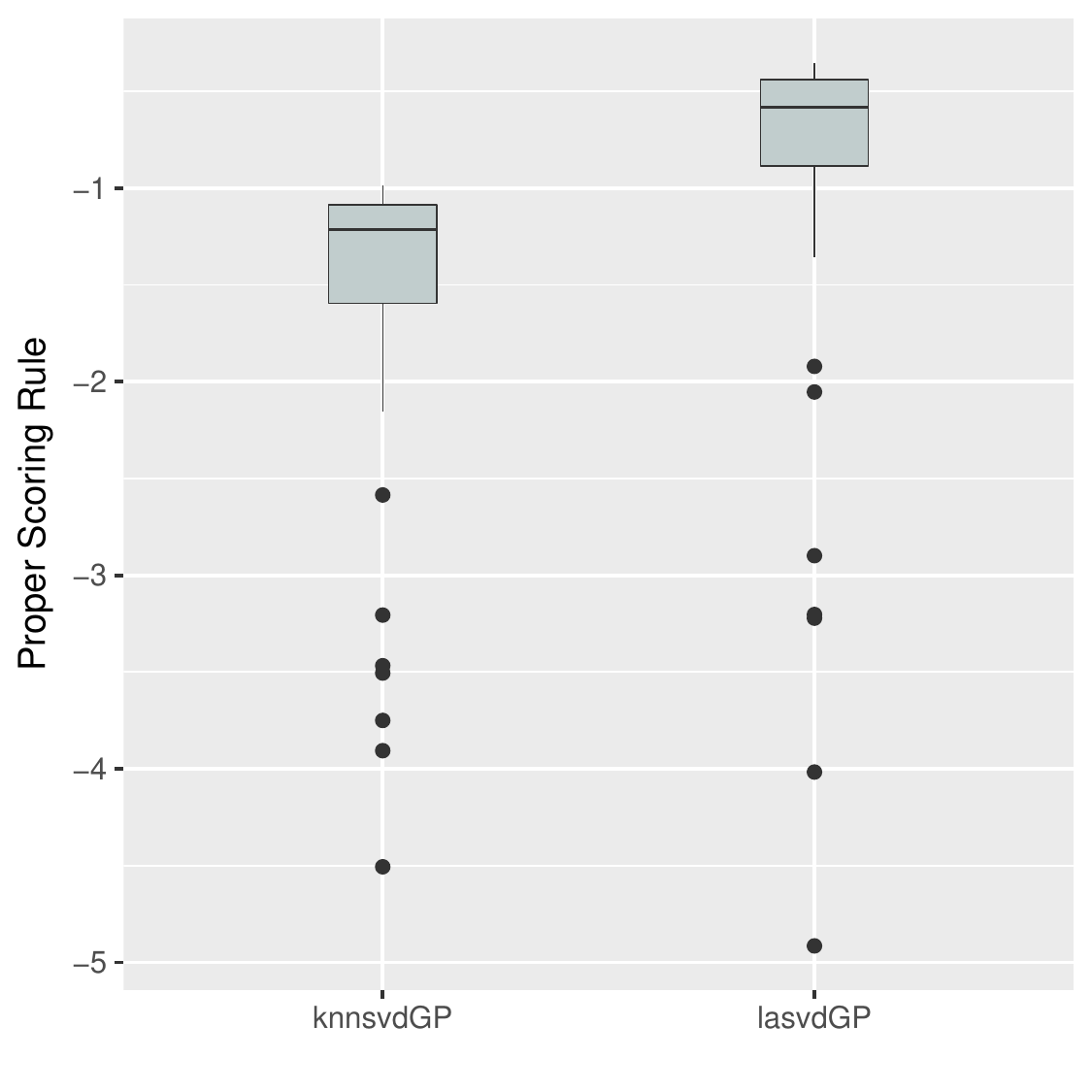}
  \caption{The boxplots of the log of mean NMSPEs (left) and mean proper scoring rule (right) for the TDB application obtained via Monte Carlo cross-validation. The proposed lasvdGP approach outperforms the knnsvdGP approach in terms of both log of mean NMSPEs and the mean proper scoring rule.}
  \label{fig:tdb_accuracy}
\end{figure}

The input variable domains are decided by expert knowledge. For convenience, we transform the inputs into 11-dimensional unit hypercube.  Given that we have a limited (data) budget from the simulator, we rely on the Monte Carlo cross-validation error alone. We had access to a data set of size 30,000 for the emulation and prediction accuracy measurements. For such a large scale dynamic computer model, svdGP is computationally infeasible. We used $n=80$ and $n_0=\ceil{n/2}$ for the proposed local SVD-based GP models. For each method, the total data was partitioned into training and test set in 4:1 ratio, and then the prediction accuracy measures were computed on the test set. Figure~\ref{fig:tdb_accuracy} shows the boxplots of the log of mean NMSPEs and mean proper scoring rule values over 50 randomly chosen Monte Carlo partitions.

\section{Concluding Remarks}
\label{sec:con-rmks}

We have proposed local approximate SVD-based GP models for large-scale dynamic computer experiments. The proposed local SVD-based GP models with the proposed neighborhood selection algorithm reduce the time complexity of the full SVD-based GP models. Though slightly more time consuming than its naive counterpart, lasvdGP has been shown to be much more accurate in prediction for both simulation examples and the real data analysis. With the assistance of parallel computation, the proposed algorithm can easily handle dynamic computer experiments with training set as large as (approx) 25,000 points, which is beyond the capacity of the full model.

There are a few remarks worth mentioning. First, in this article, we refer to large-scale dynamic computer experiments as those with a large number of inputs. This is different from the large data aspect in \cite{gu2016parallel} where the spatial-temporal applications with small run sizes (in the order of hundreds) but large numbers of time points (in the order of tens of thousands) were considered. In their application, fitting full SVD-based GP models are still computationally feasible, as the number of significant singular values might be large but the number of inputs (for $n\times n$ correlation matrix factorization) would be small.

Second, the formula (\ref{eq:l2}) does not consider the possible update of the estimated correlation parameters. If new data arrives, the empirical Bayesian estimators of correlation parameters $\btht$'s in (\ref{eq:theta}) are expected to change with the training  set. To address this issue, \citet{gramacy2015local} suggested the second order Taylor polynomial approximation.   

Third, there are possible improvements in terms of computational efficiency. In searching for neighborhood set, it has been suggested to consider more sophisticate searches such as using graphical processing units (GPUs) and approximating discrete neighborhood  searches via continuous ones along the rays emanating from each predictive point (\citet{franey2012short}; \citet{gramacy2016speeding}).
Another way to boost the computational efficiency is that instead of searching the neighborhood set for each individual point in the prediction set $\bm{X}^*$, some clustering algorithms could be performed on $\bX^*$ to divide it into groups on which the proposed neighborhood selection is executed. These are interesting topics for our future research.

\section*{Acknowledgement}
We would like to the Editor, the AE and the two referees for their valuable comments and suggestions that led to significant improvements in the article. Ranjan's research was supported by the Extra Mural Research Funding (EMR/2016/003332/MS) from the Science and Engineering Research Board, Department of Science and Technology, Government of India. Lin's research was supported by the Discovery grant from Natural Sciences and Engineering Research Council of Canada. We also thank Dr. Holger Teismann for providing the field data and outputs for the TDB model.

\section*{Supplementary Materials}

The supplementary material includes the following: 

\begin{description}

\item[Appendix:] Section~A contains the proof of Proposition  \ref{propjl}. Section~B presents two algorithms (in the formal algorithm format) for fitting local approximate SVD-based GP models (lasvdGP) described in Sections 2 and 3 of this article. Section~C summarizes simulation results for establishing the reliability of the estimated range parameters (or equivalently, the correlation parameters) for the proposed lasvdGP model fits in Examples 1 and 2. (Appendix.pdf, PDF file)

\item[Code:] \proglang{R} codes to reproduce results in the article are available in the zip file. Details can be found in the readme.txt file included. (code.zip, zipped folder)

\end{description}

\bibliographystyle{chicago}
\bibliography{lasvdgp_bib}

\newpage{}

\begin{center}
{\Large\bf Supplementary Materials}
\end{center}
 
\renewcommand{\theequation}{A.\arabic{equation}}
\section*{A. Proof Of Proposition 1}
\label{sec:derivation-19}

Following (7), the inner expectation in (10) can be written as
\begin{align}\label{eq:qfexp}
  \begin{aligned}
    &\tE\Big[\big\|\by(\bx_0)-\hat{\by}(\bx_0|\bc(\bx),\bV^{*(k)},\hbT^{(k)})\big\|^2\Big|\bc(\bx),\bV^{*(k)},\hbT^{(k)},(\hat{\sigma}^{(k)})^2\Big]\\
    =&\tr\Big(\bB^{(k)}\bm{\Lambda}\big(\bV^{*(k)}(\bx),\hbT^{(k)}\big)(\bB^{(k)})^T+(\hat{\sigma}^{(k)})^2\bI_L\Big)\\
    =&(\hat{\sigma}^{(k)})^2L+\tr\Big(\bm{\Lambda}\big(\bV^{*(k)}(\bx),\hbT^{(k)}\big)(\bB^{(k)})^T\bB^{(k)}\Big)\\
    =&(\hat{\sigma}^{(k)})^2L+\sum_{i=1}^{p_k}(d_i^{(k)})^2\hat{\sigma}^2_i\big(\bm{x}_0|\bv_i^{(k)}(\bx),\hat{\bm{\theta}}_i^{(k)}\big),
  \end{aligned}
\end{align}
where $\bV^{*(k)}(\bx)=[(\bV^{*(k)})^T,\bc(\bx)]^T$ and $d^{(k)}_i$ is the $i$th largest singular value of $\bY^{(k)}$,
\begin{align*}
\bm{\Lambda}\big(\bV^{*(k)}(\bx),\hbT^{(k)}\big)=\diag\Big(\hsq_1\big(\bx_0|\bv_1^{(k)}(\bx),\hbt^{(k)}_1\big),\dots,\hsq_{p_k}\big(\bx_0|\bv_{p_k}^{(k)}(\bx),\hbt^{(k)}_{p_k}\big)\Big),
\end{align*}
and
\begin{align*}
  \hsq_i\big(\bx_0|\bv_i^{(k)}(\bx),\hbt^{(k)}_i\big)=\frac{\rho^{(k)}_i(\bx_0,\bx)}{\alpha_i+k}\big(\beta_i+\psi_i^{(k)}(\bx)\big),
\end{align*}
for $i=1,\dots,p_k$, where  $\psi_i^{(k)}(\bx)=\bv_i^{(k)}(\bx)^T\tbK_i^{-1}(\bx)\bv_i^{(k)}(\bx)$, and $\bv_i^{(k)}(\bx)=[(\bv_i^{(k)})^T,c_i(\bx)]^T$ is the $i$th column of
$\bV^{*(k)}(\bx)$.

The first equality of (\ref{eq:qfexp}) follows from Theorem
3.2b.1 of Mathai and Provost (1992). The third equality is derived
from the column-orthogonality of $\bB^{(k)}$, i.e. $(\bB^{(k)})^T\bB^{(k)}=(\bD^{*(k)})^2$. Plugging (\ref{eq:qfexp}) into (10), we get
\begin{align*}
  J(\bx_0,\bx)&=\tE\Big[(\hat{\sigma}^{(k)})^2L+\sum_{i=1}^{p_k}(d_i^{(k)})^2\hat{\sigma}^2_i\big(\bm{x}_0|\bv_i^{(k)}(\bx),\hat{\bm{\theta}}_i^{(k)}\big)\Big|\bV^{*(k)},\hbT^{(k)},(\hat{\sigma}^{(k)})^2\Big]\\
              &=(\hat{\sigma}^{(k)})^2L+\sum_{i=1}^{p_k}(d_i^{(k)})^2\Big(\frac{\rho_i^{(k)}(\bx_0,\bx)}{\alpha_i+k}\big(\beta_i+\tE[\psi_i^{(k)}(\bx)|\bV^{*(k)},\hbT^{(k)},(\hat{\sigma}^{(k)})^2]\big)\Big)\\
              &=(\hat{\sigma}^{(k)})^2L+\sum_{i=1}^{p_k}(d_i^{(k)})^2\Big(\frac{\rho_i^{(k)}(\bx_0,\bx)}{\alpha_i+k}\big(\beta_i+\tE[\psi_i^{(k)}(\bx)|\bv^{(k)}_i,\hbt^{(k)}_i]\big)\Big)\\
              &=(\hat{\sigma}^{(k)})^2L+\sum_{i=1}^{p_k}(d_i^{(k)})^2\Big(\frac{\rho_i^{(k)}(\bx_0,\bx)}{\alpha_i+k}\big(\beta_i+\frac{\alpha_i+k}{\alpha_i+k-1}\psi_i^{(k)}\big)\Big).
\end{align*}
The second equality holds because $\rho^{(k)}_i(\bx_0,\bx)$ is
a deterministic function of $\bx_0$, $\bx$ and $\hbt_i^{(k)}$. The third
equality follows from the independence among $c_i$'s. The validity of
the fourth equality is due to Gramacy and Apley (2015).

\section*{B. Algorithms}

Algorithm~1 summarizes the key steps required for estimating the necessary parameters in the posterior predictive distribution (Equation (9) of the main article) of a full SVD-based GP model fitted to a training data of size $N$.

{\small
\begin{algorithm}[h!]
  \SetKwInOut{Input}{Input}\SetKwInOut{Output}{Output}
  \DontPrintSemicolon
  \SetKwFunction{svdGP}{svdGP}
  \SetKwFunction{buildBasis}{buildBasis}
  \SetKwFunction{inference}{inference}
  \SetKwFunction{SVD}{SVD}
  \Input{(1) Training set: $\bm{X}_{N\times q}$, (2) response matrix: $\bm{Y}_{L\times N}$, (3) threshold $\gamma$,\\ (4) prior parameters: $\balp=[\alpha_1,\dots,\alpha_p,\alpha]^T$, $\bbt=[\beta_1,\dots,\beta_p,\beta]^T$.}
  \Output{(1) Basis $\bB_{N\times p}$, (2) singular values $\bD^{*}_{p\times p}$, (3) coefficients $\bV^*$, \\ (4) correlation parameters $\hbT$, (5) variance $\hat{\sigma}^2$.}
  \nonl\hrulefill\\
  \SetKwProg{myproc}{Function}{}{}
  \myproc{\svdGP{$\bX$,$\bY$,$\balp$,$\bbt$,$\gamma$}}{

    $[\bB,\bD^*,\bV^*,p]\leftarrow$ \buildBasis{$\bY$,$\gamma$}\;
    $\bm{r}\leftarrow\text{vec}(\bY)-(\bI_N\otimes\bB)\text{vec}(\bV^{*T})$\;
$\hat{\sigma}_i^2(\bm{x}_0|\bv_i,\bm{\theta}_i)=(\beta_i+\psi_i)\Big(1-\bm{k}^T_i(\bm{x}_0)\bm{K}^{-1}_i\bm{k}_i(\bm{x}_0)\Big)/(\alpha_i+N)$, 
    $\hat{\sigma}^2\leftarrow \big(\br^T\br+\beta\big)/(NL+\alpha+2)$\;\label{alg:sigma}
    $\hbT\leftarrow$ \inference{$\bV^*$, $p$,
      $\balp$, $\bbt$}\;
    \KwRet $\bB$, $\bD^*$, $\bV^*$, $\hbT$, $\hat{\sigma}^2$
  }
    \SetKwProg{myproc}{Subroutine}{}{}
    \myproc{\buildBasis{$\bY$,$\gamma$}}{
      $[\bU,\bD,\bV]\leftarrow$\SVD{$\bY$}{\footnotesize\tcc*{perform
          SVD on matrix $\bY$.}}\label{alg:svd}
      $p\leftarrow
      \min\left\{m:\frac{\sum_{i=1}^md_i}{\sum_{i=1}^{k}d_i}>\gamma\right\}${\footnotesize \tcc*{where
        $\bD=\diag(d_1,\dots,d_N), k=\min\{N,L\}$}}
      $\bB \leftarrow \bU^*\bD^*${\footnotesize\tcc*{as in
          Section~2.1}}
      \KwRet $\bB$, $\bD^*$, $\bV^*$, $p$
    }
    \SetKwProg{myproc}{Subroutine}{}{}
    \myproc{\inference{$\bV^*$, $p$, $\balp$, $\bbt$}}{
      \For{$i\leftarrow 1$ \KwTo $p$}{\label{alg:ebi}
        $\hat{\bm{\theta}}_i\leftarrow\underset{\bm{\theta}_i}{\mathrm{argmax}}\:\pi(\bm{\theta}_i|\bv_i)${\footnotesize\tcc*{fit $p$ independent GPs by finding the MAPs}}
      }
      \KwRet $\hbT = [\hbt_1,\dots,\hbt_p]^T$
    }
\caption{SVD-based GP model}\label{algo_svdgp}
\end{algorithm}
}

\newpage

Algorithm~2 presents the steps required for fitting the proposed local approximate SVD-based GP model (lasvdGP) with the neighbourhood points selected using the $J$-criterion in Section~3.1 of the main article.

{\small
\begin{algorithm}[h]
  \SetKwInOut{Input}{Input}\SetKwInOut{Output}{Output}
  \DontPrintSemicolon
  \SetKwFunction{svdGP}{svdGP}
  \SetKwFunction{SVD}{SVD}
  \Input{(1) Training set: $\bX_{N\times q}$, (2) response matrix:
    $\bm{Y}_{L\times N}$, (3) test set $\bm{X}^*_{M\times q}$,\\ (4) neighborhood size $n$, (5) initial
    neighborhood size $n_0$, (6) threshold $\gamma$,\\ (7) prior parameters
    $\balp=[\alpha_1,\dots,\alpha_p,\alpha]^T$ and
    $\bbt=[\beta_1,\dots,\beta_p,\beta]^T$.}
  \Output{(1) The predicted mean response, and (2) the associated posterior variance in estimating $\by(\bx_0)$ for each
    $\bm{x}_0\in\bm{X}^*$.}
  \nonl\hrulefill\\
  \For{each $\bm{x}_0\in \bm{X}^*$}{\label{alg:main}
      $\bX^{(n_0)}\leftarrow\{\bx_{i}, i=1,\dots, n_0\}$\label{alg:init}
      {\footnotesize \tcc*{$n_0$ nearest neighbours of $\bx_0$ in $\bX$ as in knn}}
      $\bY^{(n_0)}\leftarrow \{y(\bx):\bx\in\bX^{(n_0)}\}$\;
      \For{$k\leftarrow n_0$ \KwTo $n-1$}{\label{alg:inner}
        $[\bB^{(k)},\bD^{*(k)},\bV^{*(k)},p_k,\hbT^k,(\hat{\sigma}^{(k)})^2,(\hat{\bm{\sigma}}^{(k)})^2]\leftarrow$\svdGP{$\bX^{(k)}$,$\bY^{(k)}$,$\balp$,$\bbt$,$\gamma$}\;
        $\bm{x}^*_{k+1}\leftarrow\underset{\bm{x}\in
          \bm{X}\backslash\bX^{(k)}}{\mathrm{argmin}}\:J(\bx_0,\bx)$\label{alg:min}\;
        $\bX^{(k+1)}\leftarrow\bX^{(k)}\cup\bm{x}_{k+1}^*$\;
        $\bY^{(k+1)}\leftarrow\bY^{(k)}\cup \by(\bx^*_{k+1})$\;
      }
      $[\bB^{(n)},\bD^{*(n)},\bV^{*(n)},p_n,\hbT^{(n)},(\hat{\sigma}^{(n)})^2,(\hat{\bm{\sigma}}^{(n)})^2]\leftarrow$
      \svdGP{$\bX^{(n)}$,$\bY^{(n)}$,$\balp$,$\bbt$,$\gamma$}\label{alg:finish}\;
      Predict $\bm{y}(\bm{x}_0)$ through $\pi(\bm{y}(\bm{x}_0)|\bV^{*(n)},\hat{\bm{\Theta}}^{(n)},(\hat{\sigma}^{(n)})^2,(\hat{\bm{\sigma}}^{(n)})^2)$
      in Eqn. (9)\label{alg:pred}\;
    }
    \caption{Proposed local SVD-based GP model}\label{algo_dyn}
\end{algorithm}
}

\section*{C. Additional Simulation Results}

We now investigate the reliability of the estimated range parameters (or equivalently, the correlation parameters) for the proposed local approximate SVD-based GP model (lasvdGP) fits in Examples 1 (Forrester et al., 2008 -- $q=3, N=10000, M=2000$, $n=40$ and $n_0=20$) and 2 (Bliznyuk et al., 2008 -- $q=5, N=10000, M=2000$, $n=50$ and $n_0=25$) of the main article.

To explain the results, recall that for each point in the test set, the SVD-based GP model fitted on the neighbourhood set is represented using a $p$-dimensional basis as in Equation (1), where $p$ is selected by the cumulative percentage criterion (Equation (2)). That is, for each test point, $p$ independent GP models for each $c_i(\textbf{x})$ are fitted in the respective neighbourhood searched.  The value of $p$ may be different for different test points. The frequency table of the number of leading basis functions for 2,000 test points for each of the two examples are displayed in Table~\ref{tab:distp}.

\begin{table}[h!]
  \centering
  \begin{tabular}{|l|llllll|l|}
    \hline
    &\multicolumn{7}{c|}{$p$}\\\cline{2-8}
    &3&4&5&6&7&8&total\\
    \hline
    Example 1& 266 & 1734 &0 & 0 & 0  & 0 & 2000\\
    Example 2& 15  & 161  &873 & 825 & 124 & 2 & 2000\\
    \hline
  \end{tabular}
  \caption{Frequency of $p$ among the 2,000 test points in Examples
    1 and 2.}
  \label{tab:distp}
\end{table}

\begin{figure}[h!]
  \centering
  \includegraphics[width=6.5in]{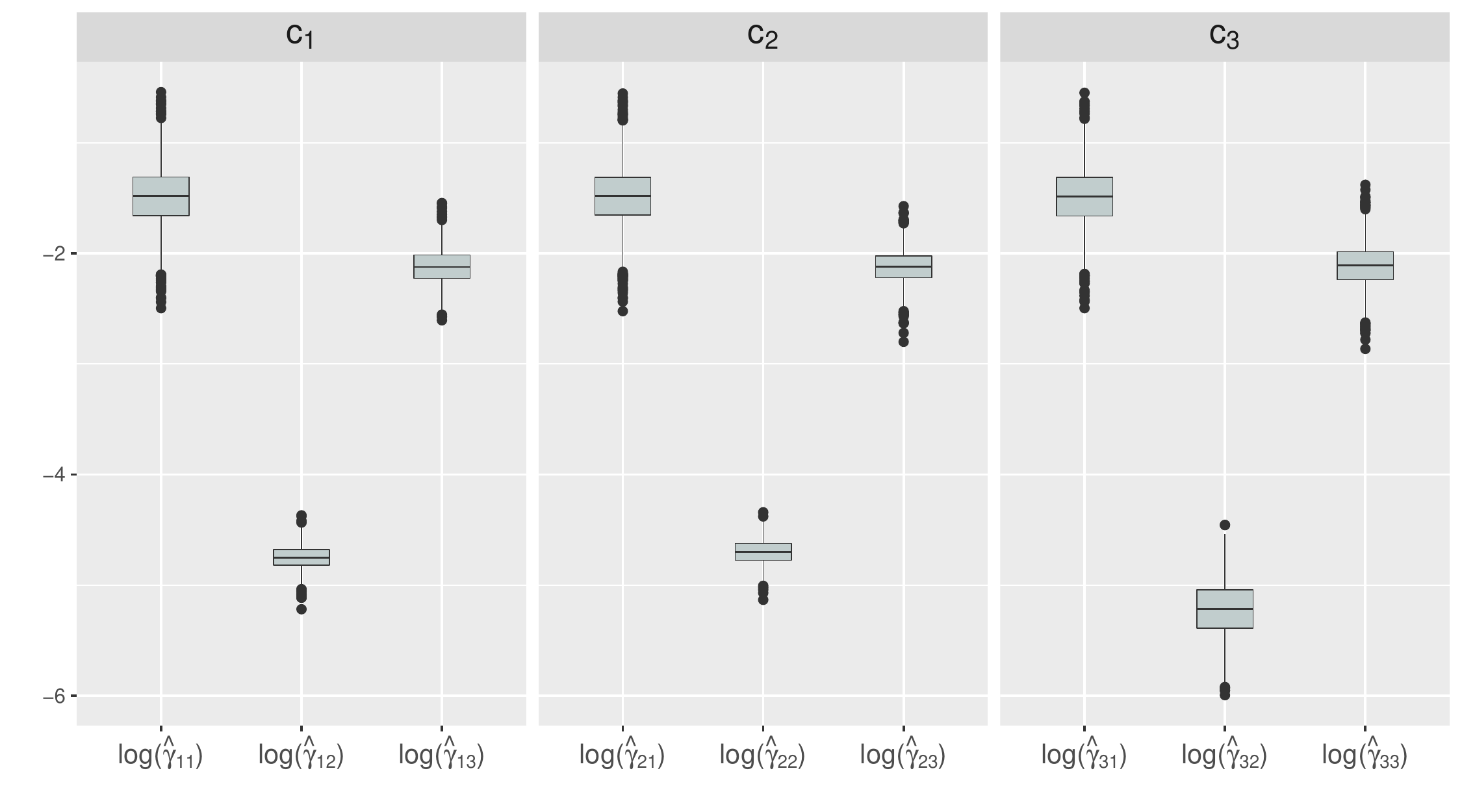}
  \caption{The boxplots of the 2,000 estimates of the log-range parameters in the GP models for  $c_1(\bx)$, $c_2(\bx)$ and $c_3(\bx)$ in Example 1 (Forrester et al., 2008). }
\label{fig:forrange}
\end{figure}

\begin{figure}[h!]
  \centering
  \includegraphics[width=6.5in]{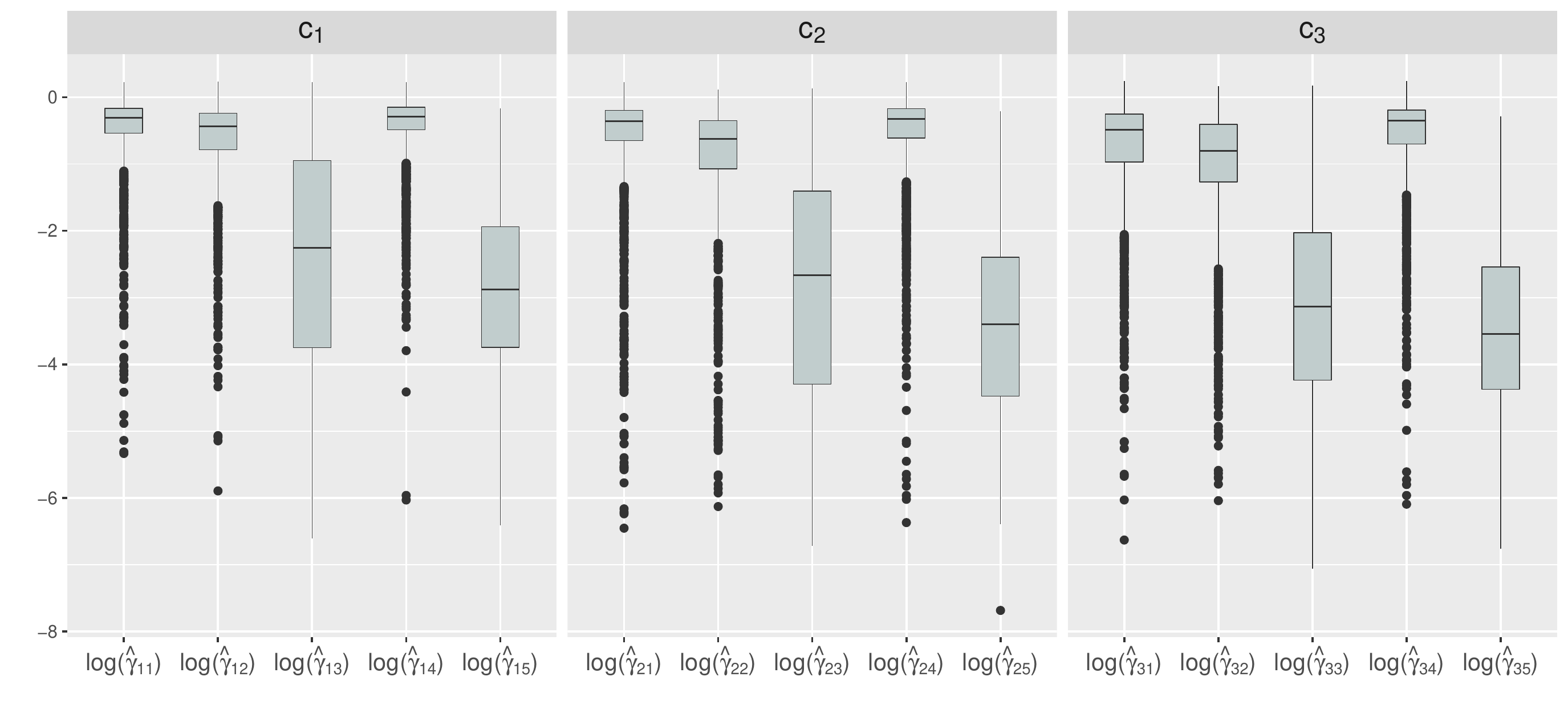}
  \caption{The boxplots of the 2,000 estimates of the log-range parameters in the GP models for  $c_1(\bx)$, $c_2(\bx)$ and $c_3(\bx)$ in Example 2 (Bliznyuk et al., 2008).  }\label{fig:envrange}
\end{figure}

For simplicity, we only report the estimated range parameters in the GP models corresponding to $c_1(\bx)$, $c_2(\bx)$ and $c_3(\bx)$ from the final fits, i.e., after $n-n_0$ follow-up points were added. Figures~\ref{fig:forrange} and \ref{fig:envrange} display the boxplots of those 2,000 estimates for each range parameter in Examples 1 and 2, respectively.

\end{document}